\PassOptionsToPackage{unicode}{hyperref}
\PassOptionsToPackage{hyphens}{url}
\documentclass[
  10pt,
  twocolumn]{article}
\usepackage{xcolor}
\usepackage{amsmath,amssymb}
\usepackage{iftex}
\ifPDFTeX
  \usepackage[T1]{fontenc}
  \usepackage[utf8]{inputenc}
  \usepackage{textcomp} 
\else 
  \usepackage{unicode-math} 
  \defaultfontfeatures{Scale=MatchLowercase}
  \defaultfontfeatures[\rmfamily]{Ligatures=TeX,Scale=1}
\fi
\usepackage{lmodern}
\ifPDFTeX\else
\fi
\IfFileExists{upquote.sty}{\usepackage{upquote}}{}
\IfFileExists{microtype.sty}{
  \usepackage[]{microtype}
  \UseMicrotypeSet[protrusion]{basicmath} 
}{}
\makeatletter
\@ifundefined{KOMAClassName}{
  \IfFileExists{parskip.sty}{%
    \usepackage{parskip}
  }{
    \setlength{\parindent}{0pt}
    \setlength{\parskip}{6pt plus 2pt minus 1pt}}
}{
  \KOMAoptions{parskip=half}}
\makeatother
\providecommand{\tightlist}{%
  \setlength{\itemsep}{0pt}\setlength{\parskip}{0pt}}
\usepackage[numbers,sort&compress]{natbib}
\usepackage{times}
\usepackage{booktabs,array,graphicx}
\usepackage[letterpaper,left=0.75in,right=0.75in,top=1in,bottom=1in,columnsep=0.33in]{geometry}
\usepackage[font=small,labelfont=bf,labelsep=period]{caption}
\usepackage{bookmark}
\IfFileExists{xurl.sty}{\usepackage{xurl}}{} 
\hypersetup{
  pdftitle={ModelLess: A Serverless System for Multi-Tier LLM Inference on Shared GPUs},
  pdfauthor={Ding Shi; Wei Wu; Jiayuan Chen; Chenghu Yang},
  hidelinks,
  pdfcreator={LaTeX via pandoc}}

\title{ModelLess: A Serverless System for Multi-Tier LLM Inference on
Shared GPUs}
\author{Ding Shi \qquad Wei Wu \qquad Jiayuan Chen \qquad Chenghu Yang\\[0.4em]{\large Ant Group}}
\date{}

\begin{document}
\maketitle
\begin{abstract}
LLM inference platforms sell online, asynchronous, batch and rollout
products on GPUs provisioned per product, so a service's cards sit idle
between bursts while other work waits. We present ModelLess, a
serverless system for LLM inference. Users declare model services and
the platform owns the instances. An idle service scales to zero: its
instances go to sleep with their weights in host memory (release about 1
s, wake about 4 s for GLM-5.2 at TP8, against a cold start of over four
minutes). Billing follows the time a service holds GPUs. The GPUs that
sleeping instances free are lent to lower tiers. Cluster managers
preempt low-priority tasks and treat eviction as all-or-nothing, but an
LLM instance's eviction cost depends on which of its weights, execution
state, KV cache and in-flight requests survive. ModelLess therefore
gives each tier (online, job, flex) eviction semantics that state what
it may lose when preempted, and runs a preemption protocol that closes
admission before it releases memory, re-enqueues a request durably
before it aborts it, and fails closed. Four components (entry,
dispatcher, node agent, SGLang plugin) enforce it with sleep/wake,
pause/resume of in-flight requests for flex, and a weight-copy path
(follow/seal) for RL rollout.

Replaying a week of traces from 1,100 H20 cards serving long-tail online
models shows that about 38\% of GPU-time (417 equivalent GPUs) could be
lent under a coarse latency target with one fixed idle threshold. This
is an idealized upper bound, and it holds only if recovery takes
seconds: with a 240 s cold start under 8\% remains even with per-service
thresholds. On one node, a single four-hour run per policy shows that
pausing yields 22\% more borrower tokens per held GPU sample than
abort-and-requeue, with the same median online latency. The job tier and
follow/seal are not yet evaluated end to end.
\end{abstract}

\section{Introduction}\label{introduction}

Inference platforms for large language models sell several products at
once: online endpoints with latency objectives, asynchronous requests,
batch jobs and RL rollout. The products differ by orders of magnitude in
how long a request may wait, yet GPUs are provisioned per product. Each
online service holds cards for its peak, and rollout and batch work run
on separate capacity. Cards held by an idle online service cannot serve
a waiting batch job, and the batch pool must be sized for its own peak.

ModelLess takes the serverless view. Users declare model services, not
instances, and the platform owns instances, placement and scaling. A
service with no traffic scales to zero: its instances sleep and hold no
GPU. Billing follows the time a service holds GPUs, so a sleeping
instance incurs no GPU charge. Scale-to-zero has a catch for LLMs: a
cold start takes minutes. ModelLess keeps the weights in host memory so
that an instance wakes in seconds, and it gives the GPUs that sleeping
instances free to lower tiers.

Cluster managers preempt low-priority work to free capacity, as Borg
does with its non-production tasks \citep{verma2015borg}, but they treat
a task as opaque: it is running or it is evicted. A model instance is
not opaque. Its expensive state (weights and compiled execution state)
can be parked in host memory and restored in seconds, and its cheap
state (KV cache and in-flight requests) can be checkpointed. So eviction
cost is a design choice. Cluster managers kill. Engine-level work (sleep
mode, InferCept, ConServe) chooses what one engine keeps; none of it
hands the freed GPU to another model. ModelLess makes retention a
cluster-level property of each tier, and hands the GPU, weights
included, to another model without losing the requests in flight.

We first ask whether the idle capacity is real, using a week of traces
from 1,100 H20 cards serving long-tail online models (§2). Services are
idle most of the time, as other studies of long-tail and general serving
clusters also report \citep{xiang2025aegaeon, yu2025prism, gao2026rose},
and LLM request streams are bursty
\citep{wang2025burstgpt, xiang2026servegen}. Most idle windows are
short, but most idle GPU-time lies in windows of an hour or longer.
Under a coarse latency target, 38\% of GPU-time can be lent with one
fixed idle threshold and 46\% with per-service thresholds (§7.1). The
share depends on recovery latency. A cold start of tens of seconds to
minutes
\citep{kabakibo2026breakingice, fu2024serverlessllm, zhang2025blitzscale}
leaves only a small fraction of GPU-time reclaimable, while a recovery
of a few seconds keeps most of it. The stricter the latency promise, the
more the speed of recovery matters.

Turning this capacity into work for lower tiers requires four things.
First, the handover must take seconds, but rebuilding a model instance
takes minutes (§2.5). Second, the borrower is preempted often and its
windows are shorter than its jobs: a batch request may run for minutes,
while the owner returns within a minute or two. Preempting by
abort-and-requeue discards the decode progress and reruns the request;
preempting by recompute continuation keeps the generated tokens but
prefills the whole context again. Third, the handover must keep weights
and execution state, which take minutes to rebuild, and keep KV cache
and in-flight requests only where a tier needs them. Fourth, the
borrower's weights change while a task runs, because a rollout tier
serves the policy being trained
\citep{sheng2025hybridflow, fu2025areal, wang2025roll}.

We address these in ModelLess, a serverless system in which the model
instance is the unit of scheduling and preemption. It rests on four
design principles.

\begin{enumerate}
\def\labelenumi{(\arabic{enumi})}
\tightlist
\item
  \emph{Schedule the instance, not the container.} An instance carries
  execution, weight, KV and request state, and every handover is a
  choice of which of these to keep (§4). An object chain separates what
  is declared from what runs, so that one base runtime can host
  workloads of different shapes without a restart and the tier is not
  coupled to the engine (§3.1). (2) \emph{Give each tier its own
  eviction semantics, matched to what its clients tolerate.} An online
  request in flight tolerates no stall, so a lower tier never preempts
  it. An online service idle for \ensuremath{\theta} seconds may be
  parked, and its first request then pays the recovery latency. A job is
  issued for a trainer that waits on it, so its requests on engines
  above \texttt{min} are migrated and run slower, but they are not
  paused. Flex clients submit to a queue and poll, so a flex request may
  be fenced and requeued, or paused where the engine supports it (§4.2).
  (3) \emph{Protect the owner by exclusive handover.} One engine is
  active on a card group at a time, and a handover gives the whole group
  back (§6.2). GPUs cannot be partitioned as finely as CPUs, and large
  models fill a card group. (4) \emph{Safety comes from ordering and
  failing closed.} The preempted engine's admission gate closes before
  its memory is released. An attempt is re-enqueued before it is
  aborted. A failed pause leaves the engine in a recovery-required state
  with admission closed (§5.2, §6.2).
\end{enumerate}

Four components implement these principles (§3.3). \emph{modelentry}
holds the client connection and fronts the durable queue.
\emph{modeldispatcher} makes every placement decision. \emph{modellet}
owns a node's cards and the engine lifecycle. An \emph{engine plugin}
inside SGLang carries out sleep, wake and pause.

The components act on the instance with three operations and two
request-level preemption paths, requeue and recompute continuation.
\emph{Sleep/wake} releases GPU memory while keeping weights in host
memory \citep{vllm-sleep-mode, torch-memory-saver}, so recovery takes
seconds. \emph{Pause/resume} checkpoints in-flight requests, with their
KV pages and sampling counters, to host memory and restores them after
the next wake, so a preempted borrower continues where it stopped. For
the flex tier it is an alternative to \emph{requeue}. The job tier uses
\emph{recompute continuation}, which resubmits the prompt plus the
tokens generated so far. For the fourth requirement the design has
\emph{follow} and \emph{seal}. A new engine copies its weights from a
running engine of the same weight version. An engine becomes such a
source only after it has produced a successful generation with those
weights (§5.3).

The platform's commercial products map onto the three tiers: online
products belong to the online tier, rollout to the job tier, and
asynchronous and batch products to the flex tier (§4.2). The two
workloads that shaped the design are long-tail online services, whose
idle capacity we measure, and RL rollout, whose engines serve weights
that change every training step.

On one eight-GPU node, a four-hour comparison of two preemption policies
that differ in a single switch measures what pausing is worth against
abort-and-requeue (§7.3). With the same binary, online trace and
borrower workload, pausing gives the borrower 22\% more tokens per held
GPU sample (once dead time from engine failures in the requeue run is
excluded; 25\% without the exclusion). The median online latency does
not change (176 vs 170 ms), and the tail is set by the handover in both
runs (§7.3, §8).

\textbf{Contributions.}

\begin{enumerate}
\def\labelenumi{\arabic{enumi}.}
\tightlist
\item
  A serverless resource model: users declare model services, the
  platform schedules instances, and each tier states which of four
  layers of instance state it may lose when preempted (§3, §4).
\item
  The design and implementation of ModelLess on SGLang: four components
  (modelentry, modeldispatcher, modellet and an engine plugin), handover
  operations that keep the expensive state layers (sleep/wake,
  pause/resume that fails closed, and follow/seal), and a preemption
  protocol with safety invariants and a fail-closed engine state machine
  (§3--§6).
\item
  A measurement of idle capacity in a production inference fleet at
  one-second resolution, with a bound on what can be lent as a function
  of recovery latency and SLO strictness, which fixes the requirements
  of the design (§2, §2.5, §7.1).
\item
  Measured handover costs and a controlled four-hour comparison on one
  node of pausing against abort-and-requeue, with a second-level account
  of where pool time goes, and the limits of the evidence (§7, §8).
\end{enumerate}

§2 describes the tiers, the idle capacity and the requirements they
impose. §3 presents the resource model and architecture, §4 instance
state and tier semantics, §5 the handover operations, and §6 the control
plane and its safety invariants. §7 is the evaluation and §8 discusses
limitations and future work. §9 relates ModelLess to prior work on LLM
serving, serverless keep-alive and RL rollout.

\section{Motivation: Tiers and Idle
Capacity}\label{motivation-tiers-and-idle-capacity}

\subsection{Tiered Products and Static
Partitions}\label{tiered-products-and-static-partitions}

A production inference platform sells more than one kind of service.
Customers may call online endpoints under latency objectives, submit
asynchronous requests, or queue batch jobs. These products differ in the
time scale of their service-level objectives (SLOs) and in how much
interruption they tolerate. We measure against two online SLOs that
illustrate the spread: a stricter SLO of a median time-to-first-token
(TTFT) below 2 s for inputs under 4K tokens with 99.5\% availability,
and a standard SLO of a median TTFT below 4 s with roughly 95\%
availability. Asynchronous and batch products accept minutes to hours.

Resources, however, are provisioned per product. Each online service
holds its own GPUs for its peak, and the asynchronous and batch products
run on separate capacity. The result is a pool that is expensive to size
and hard to rebalance, as production model-serving clusters also report
\citep{jaiswal2025sageserve, xiang2025aegaeon}. A GPU held by an idle
online service cannot serve a batch job that is waiting elsewhere. This
section measures how much capacity is idle; §2.5 states what a handover
must satisfy.

\subsection{Fleet and Observations}\label{fleet-and-observations}

We study the long tail of online services on H20 GPUs. A service enters
the population if, on a single reference day, its request rate is at
most 0.05 QPS, it has at most ten instances and ten cards, it has
existed for more than a day, and it runs on H20-class cards. We then
track the selected services over time. Some of them scale up during
tracking; in the regular week used below, about half of the services
have a mean rate above 0.05 QPS, and the busiest reaches roughly 100
QPS. We keep them in the population because the question is what happens
to a tracked fleet, not to a frozen sample.

The population contains about 1,000 services using about 4,000 cards
(Figure 1). All inference traffic enters through a single gateway, so we
observe arrivals only for services that receive gateway traffic. The
population splits into two groups. \emph{Services with traffic}: 243
services (1,182 cards) have valid inference traffic in at least one of
our two observation weeks. \emph{Services without traffic}: the
remaining services (about 3,000 cards) have no valid inference traffic;
their request rate is recorded as zero, and their GPU utilization comes
from health checks, not inference. We treat the two groups separately
throughout, because they raise different questions. For services with
traffic we must ask whether idle time can be lent without violating an
SLO. For services without traffic the SLO question does not arise, but a
health check that wakes the engine would still prevent reclamation.

\begin{figure}[t]\centering\includegraphics[width=\columnwidth]{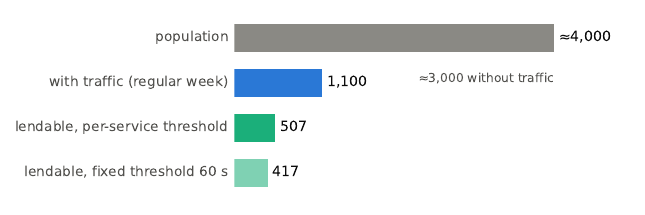}\caption{From the roughly 4,000 cards of the population to the cards that sleep
and wake can lend: 1,100 cards have traffic in the regular week, and on
average 507 (per-service thresholds) or 417 (fixed threshold) are
lendable at any moment. About 3,000 cards have no valid inference
traffic, and a few more had traffic only in the holiday week.}\end{figure}

All replay results use services with traffic, over two weeks. The
regular week covers 223 of these services on 1,100 cards; this is the
analysis population behind every fleet-level figure in the paper
(including the 1,100 cards of the abstract) unless a figure is marked as
the holiday week. The National Day holiday week covers all 243 services
on 1,182 cards. The counts differ because services come online and go
offline and some cannot be retrieved in every week; each statistic uses
the denominator of its own week. The services without traffic are the
plainest case for scale-to-zero. The platform keeps them deployed for
rollbacks and later tests, and their health checks wake the engine, so
their roughly 3,000 cards are in neither denominator (§2.4). Request
arrivals have second-level resolution.

\subsection{Idle Structure}\label{idle-structure}

Idleness is common, but it is not smooth, as in other bursty LLM
workloads \citep{wang2025burstgpt, xiang2026servegen}. Between requests,
most long-tail services leave their GPUs unused for long stretches,
interleaved with bursts. Our replay (§7.1) parks a service when no
request has arrived for \ensuremath{\theta} seconds, a keep-alive-style
idle threshold \citep{shahrad2020serverless}, and releases the GPUs
until the next arrival. With \ensuremath{\theta} = 60 s, a 5 s recovery,
and no SLO constraint, 47.8\% of the regular week's GPU-time is parked;
with \ensuremath{\theta} = 30 s the figure is 52.6\%.

Table 1 and Figure 2 show that parked segments are short by count but
long by weight. At \ensuremath{\theta} = 60 s the median segment lasts
only 62 s, yet 59\% of the parked GPU-seconds fall in segments longer
than one hour, and the median parked card-second sits in a segment of
2.1 h. The cost of this structure is churn: at \ensuremath{\theta} = 60
s a card is woken by its owner about 99 times per day. A borrower
therefore receives most of its GPU-time in long windows, but most of its
windows are short (median 62 s); windows under ten minutes carry 21\% of
the GPU-time.

\begin{table*}[t]
\caption{Parked segments by idle threshold \ensuremath{\theta} (regular week, 5 s
recovery; gross, no SLO gate). ``Owner wakeups'' counts arrivals that
end a parked segment.}\centering\footnotesize
\begin{tabular}{>{\raggedright\arraybackslash}p{\dimexpr(\textwidth-14\tabcolsep)*49/1000\relax}>{\raggedright\arraybackslash}p{\dimexpr(\textwidth-14\tabcolsep)*79/1000\relax}>{\raggedright\arraybackslash}p{\dimexpr(\textwidth-14\tabcolsep)*159/1000\relax}>{\raggedright\arraybackslash}p{\dimexpr(\textwidth-14\tabcolsep)*153/1000\relax}>{\raggedright\arraybackslash}p{\dimexpr(\textwidth-14\tabcolsep)*190/1000\relax}>{\raggedright\arraybackslash}p{\dimexpr(\textwidth-14\tabcolsep)*196/1000\relax}>{\raggedright\arraybackslash}p{\dimexpr(\textwidth-14\tabcolsep)*171/1000\relax}}\toprule
\ensuremath{\theta} (s) & Parked share \ensuremath{\uparrow} & Owner wakeups / card / day & Median segment (by count) & Median segment (by card-second) & GPU-seconds in segments \textless{} 10 min & GPU-seconds in segments \ensuremath{\geq} 1 h \\
\midrule
30 & 52.6\% & 194 & 29 s & 1.4 h & 27\% & 54\% \\
60 & 47.8\% & 99 & 62 s & 2.1 h & 21\% & 59\% \\
300 & 37.1\% & 16 & 287 s & 4.6 h & 6\% & 74\% \\
1800 & 26.7\% & 2.8 & 1,739 s & 10 h & 1\% & 90\% \\
\bottomrule\end{tabular}
\end{table*}

\begin{figure}[t]\centering\includegraphics[width=\columnwidth]{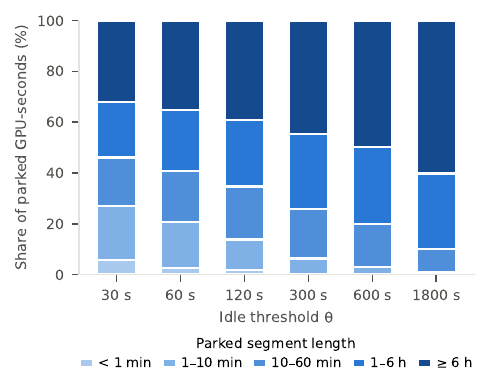}\caption{Parked GPU-seconds by segment length for each idle threshold
\ensuremath{\theta} (regular week, C = 5 s). Most GPU-time is in
segments longer than an hour, even though most segments are short.}\end{figure}

\subsection{Reclaimable Capacity}\label{reclaimable-capacity}

Not all parked time can be lent. A parked service must still honor its
own SLO when its next request arrives, and a borrower needs time to load
a model and time to yield. We therefore define \emph{reclaimable
capacity} as the GPU-time that can be parked while each service's SLO
gate holds, and \emph{net reclaimable capacity} as that quantity after
subtracting the borrower's load and yield time and discarding segments
too short to be useful.

§7.1 defines the gates. With the median-and-timeout gate (median TTFT
and a 90-second timeout rate of at most 0.5\%), the 5 s recovery, and
each service's best threshold, the gross reclaimable share of the
regular week is 47.7\%. Subtracting a 5 s borrower load, a 3 s yield,
and a 10 s minimum useful segment leaves 46.1\%, which is 507 GPUs of
equivalent steady capacity out of 1,100 cards. A single fixed
\ensuremath{\theta} = 60 s, which does not rely on per-service tuning,
gives 417 GPUs (38\%). In the holiday week the same procedure gives
54.5\%, or 644 GPUs. Overheads matter little: raising the borrower load
from 2 s to 30 s (yield fixed at 3 s) reduces the figure only from 512
to 474 GPUs, because the segments that overheads remove are short. What
reduces capacity is the SLO gate and the recovery latency, which §7.1
quantifies.

These figures are upper bounds. They treat a service as one lendable
unit and assume that its pods are idle together, ignore node,
tensor-parallel and fragmentation constraints, assume a borrower with
unlimited demand, and choose \ensuremath{\theta} after the fact.
Shielding health checks from the engine could add up to about 3,000 more
cards; we did not measure this.

\subsection{Handover Requirements}\label{handover-requirements}

A handover has an \emph{owner}, the online service that holds a card
group, and a \emph{borrower}, a lower-tier workload that uses it while
the owner is idle. \emph{Lending} begins after the owner has received no
request for \ensuremath{\theta} seconds, the \emph{idle threshold}: the
owner's instance is parked (put to sleep) and the borrower's instance
wakes. \emph{Preemption} of the borrower begins when the owner's next
request arrives: the borrower must leave and the owner's instance must
return. Three durations describe it. The borrower's \emph{yield time} is
the time to vacate the cards. The owner's \emph{wake time} is the time
to bring its instance back. The borrower's \emph{load time} is what it
needs before it can work in a window. We write C for the \emph{recovery
latency}, the extra wait of an owner request that triggers a preemption.
C is the wake time of the owner's instance, and the borrower's yield
time adds to it. A window of length L gives the borrower at most L minus
its load and yield times of useful work. C = 5 s corresponds to
sleep/wake with weights in host memory, and C = 240 s to a recovery that
reads weights from disk, comparable to a cold start. The trace study
counts only the wake time in C and charges the yield to the borrower, so
its owner-side gates omit the yield. C = 5 s understates the tail: the
end-to-end run of §7.3 measured a 99th-percentile online TTFT of 10--11
s (§7.3.2). Four requirements follow from the measurements above.

\textbf{Recovery must take seconds, and how many seconds depends on the
SLO.} A restart from disk (C = 240 s) breaks the SLO of most services
that have traffic. Others also report cold starts of tens of seconds to
minutes
\citep{kabakibo2026breakingice, fu2024serverlessllm, zhang2025blitzscale}.
At any \ensuremath{\theta} from 30 s to 300 s, 69--85\% of these
services exceed the 0.5\% timeout threshold. Only 8\% of GPU-time is
then reclaimable under the median-and-timeout gate (Table 8, §7.1). An
assumed restart from weights preloaded in host memory (C = 60 s) keeps
42.5\% under the median-and-timeout gate, against 47.7\% for C = 5 s.
Under a mean-TTFT gate it keeps 18.1\% against 46.5\%, and under tail
gates it loses about a quarter. The recovery budget is a few seconds for
services that promise tail latency and up to about a minute for services
with only a median and a timeout. The borrower's yield time adds to the
owner's wait, so it must be small too; in the four-hour comparison of
§7.3 it took a few seconds.

\textbf{Borrowers get long windows in aggregate but are preempted
often.} At \ensuremath{\theta} = 60 s, 59\% of parked GPU-seconds lie in
segments longer than an hour. Yet the median segment lasts about a
minute, and an owner wakes a card about 99 times a day. Most of a
borrower's GPU-time comes in windows that last hours, so a model load of
seconds is negligible: net reclaimable capacity falls by under 10\% when
the load time grows from 2 s to 30 s (§2.4). What each preemption costs
the borrower then decides how much of the borrowed time becomes work. A
request that is re-run from scratch may never finish inside a window; a
recomputed one costs time in proportion to its context; a kept one costs
a few seconds per preemption. The choice depends on what the borrower's
clients can tolerate, so the eviction semantics differ by tier (§4.2).

\textbf{Keep what is expensive to rebuild.} Weights and the execution
environment cost minutes to rebuild, so a handover keeps them: sleep
keeps the weights in host memory, and every runtime is prepared once
before it serves. KV cache and in-flight requests are cheap to keep but
lost when GPU memory is released. A handover keeps them only where the
workload needs them, and §4 turns that need into per-tier eviction
semantics.

\textbf{Weights change during a task.} RL rollout does not hold fixed
weights. Each training step produces a new weight version that every
rollout engine must serve before it generates, and engines are added and
removed during a step. Loading each version from storage would put the
tens of gigabytes of a large model on the critical path of every step, a
known cost in RL systems
\citep{sheng2025hybridflow, fu2025areal, wang2025roll}. The trainer
pushes weights into the engines, so the only copies live in engines. A
new engine therefore follows one that already holds the step's weights
and copies them from its read-only snapshot in host memory (§5.3).

Table 2 summarizes the requirements and what answers them.

\begin{table*}[t]
\caption{Requirements and what answers them.}\centering\footnotesize
\begin{tabular}{>{\raggedright\arraybackslash}p{\dimexpr(\textwidth-6\tabcolsep)*333/1000\relax}>{\raggedright\arraybackslash}p{\dimexpr(\textwidth-6\tabcolsep)*333/1000\relax}>{\raggedright\arraybackslash}p{\dimexpr(\textwidth-6\tabcolsep)*333/1000\relax}}\toprule
Requirement & Evidence & Answer (section) \\
\midrule
Recovery in seconds, with a small yield time & Table 8: cold start is outside the SLO; C = 5 s versus 60 s matters under strict gates & Sleep/wake with weights in host memory (§5.1); yield through pause or drain (§5.2) \\
Borrowers are preempted often, windows are long in aggregate & Fig. 2; Table 1 & Per-tier eviction semantics (§4.2, §5.4) \\
Keep what is costly to rebuild or recompute & Weights take minutes to reload; recomputing a long context takes tens of seconds (Table 11) & Sleep/wake keeps weights (§5.1); KV and requests are kept per tier: pause for flex (§5.2), requeue or recompute continuation otherwise (§5.4) \\
Weights evolve during a task & Design requirement of RL rollout & Direct push, seal, follow (§5.3) \\
\bottomrule\end{tabular}
\end{table*}

\subsection{Existing Mechanisms}\label{existing-mechanisms}

Each existing mechanism loses at least one of the four layers of
instance state (§4.1). \emph{Cluster-manager priority and preemption},
with a grace period for graceful termination, evicts the pod. The
process dies, so the execution and weight layers are lost and the
restart costs minutes, about 250 s for our largest model (§7.2). A grace
period only drains requests. A running pod's GPUs cannot be given to
another pod while its process lives, and a controller that drove sleep
from outside would lack the per-request identity that the ordering rules
of §6.2 need. \emph{Plain scale-to-zero} (Knative, KEDA) deletes the
pod, so the next request pays a full cold start, which Table 8 shows to
be outside the SLO of most services. ModelLess keeps scale-to-zero and
makes it cheap with sleep and wake. \emph{Faster cold starts}
(ServerlessLLM, BlitzScale) narrow the gap: ServerlessLLM, for example,
loads weights from a DRAM or SSD tier. A restart of any length still
loses the execution state and the KV and request layers. Table 8 shows
what a restart of about a minute would keep under each gate.
\emph{Engine sleep alone} \citep{vllm-sleep-mode, torch-memory-saver}
keeps the weights but drops the prefix cache and says nothing about
in-flight requests, tiers or the order in which a handover must happen.

Mechanisms that share or pack GPUs answer a different question.
\emph{Co-serving} (ROSE, Prism) keeps two models on the same GPUs. It
takes device memory from the online engine's KV cache: a 373 GB borrower
takes about 47 GB on each of eight 141 GB cards, a third of the memory.
GPUs also cannot yet be partitioned finely enough to isolate the two, so
we hand over whole card groups. \emph{Consolidating} long-tail services
onto fewer cards (Aegaeon, Prism) changes how many cards a service needs
but does not by itself serve the lower tiers; the windows between bursts
on the cards that remain are what lending uses. Our fleet shows how much
remains. Of about 4,000 cards, about 3,000 belong to services with no
traffic at all; consolidation would address them and we leave them out.
On the 1,100 cards of the services with traffic in the regular week,
about 507 are lendable at any moment on average and at least 342 in 95\%
of minutes (§7.1). Finally, \emph{request-level KV policies} such as
InferCept's decide what an engine does with a request it keeps; they
keep the GPU, whereas lending hands it to another model.

\section{Resource Model and
Architecture}\label{resource-model-and-architecture}

\subsection{Objects and Identities}\label{objects-and-identities}

ModelLess exposes six objects in a chain. An \emph{environment} is the
base runtime environment and node information. A \emph{catalog} entry
holds the launch parameters and base weights of one base model. A
\emph{base runtime} (or a set of them) is a catalog entry realized on
hardware, and a \emph{runtime} (or set) is a base runtime that can
serve. A \emph{model service} is an externally served model with its
capacity, and a \emph{request} is one call to it. The chain is a
decoupling across lifetimes, and it is one of the central design
choices. A base runtime---the engine process, compiled kernels, CUDA
graphs and base-model layout---lives for hours or days, whereas an RL
weight revision lives for a single training step (minutes) and a request
lives for seconds to minutes. Traditional serving replicas weld these
lifetimes together, forcing a full engine restart whenever weights or
serving shapes change. ModelLess keeps what is declared (the catalog
entry and the model service) apart from what runs (base runtimes and
runtimes). A base runtime holds the execution state and base weights of
a catalog entry and carries no priority and no workload shape; a runtime
binds a base runtime to a model service and a weight version. Workloads
of different shapes and concurrency, online, job and flex, are services
above the chain, and any base runtime beneath it can host any of them.
Moving a card group between services or advancing an RL rollout across
successive weight revisions is therefore a rebinding on the same base
runtime rather than a process restart; the dispatcher can rebuild a
runtime without changing what clients see, and the tier a request
belongs to is never baked into the engine process.

Identity is kept apart from location. A runtime is served by one engine
process, which we also call an instance, and is identified by a stable
identifier; its endpoint is only where it can currently be reached and
may change after a rebuild, so no component uses an endpoint as
identity. A catalog entry states which engines can host it as a
capability selector over labels and versions that each modellet reports,
much as a node advertises its kernel version to a cluster scheduler: the
dispatcher selects a runtime only if the reporting node satisfies the
selector. A logical weight identifier names a weight version, an empty
identifier means the base model, and a separate field gives its physical
source.

\subsection{API Definition}\label{api-definition}

Users declare and call; they do not place engines. A platform
administrator creates catalog entries, and a user creates a \emph{model
service} that names a catalog entry, optionally a weight source, a
default tier and the set of tiers its requests may use, and a replica
range (\texttt{min} and \texttt{max}). \texttt{min} is the number of
runtimes kept ready, below which the service is unfulfilled, and it may
be 0. Capacity entitlement is decoupled from physical GPU residency: a
ready runtime may be asleep and hold no GPU memory while still counting
toward \texttt{min} (and incurring no GPU charge), so reserved capacity
is guaranteed logically and materialized physically only on demand.
\texttt{max} is a bound on on-demand growth, not a promise that it is
reached at creation. The weight source is \texttt{local} (a path),
\texttt{huggingface} (a repository) or \texttt{nccl}; the last is
allowed only for a service, never for a catalog entry, because a catalog
entry must be able to cold-start its base model. For an \texttt{nccl}
service ModelLess starts the runtimes with the base weights and returns
their endpoints, rank offsets and world size, and the trainer forms its
own communication group and pushes the weights (§3.4). Changing the
physical source behind a non-empty weight identifier requires a new
service, so that a running logical weight is never silently replaced.

A request is an ordinary OpenAI-, SGLang- or vLLM-style call. Its
\texttt{model} field names the service, and it may add a weight
identifier, which must match the service's binding, and a tier, which
must be among the service's allowed tiers and otherwise defaults to the
service's tier. A request never carries a catalog, runtime or endpoint
identifier; placement is the dispatcher's. Runtimes and runtime sets are
created and sized by ModelLess alone: the replica count of a runtime set
is written only by the service's controller, as in a Kubernetes
ReplicaSet, and operators may delete a set but ModelLess rebuilds it
while its service lives. Compatibility is strict: if no reporting node
satisfies a catalog entry's selector, the service stays partial or is
rejected, and it is never downgraded to an incompatible engine.

\subsection{Components}\label{components}

ModelLess sits between the inference engine and the business gateway,
and four components divide the work (Table 3, Figure 3). It is not an
inference engine, not a gateway (it performs no authentication, quota or
billing), and not a Kubernetes controller; Kubernetes only starts the
components initially. Billing is outside ModelLess and charges for the
time a service holds GPUs. An asleep instance holds none and is not
billed, even when it counts toward \texttt{min}.

\begin{table}[t]
\caption{ModelLess components.}\centering\footnotesize
\begin{tabular}{>{\raggedright\arraybackslash}p{\dimexpr(\columnwidth-6\tabcolsep)*265/1000\relax}>{\raggedright\arraybackslash}p{\dimexpr(\columnwidth-6\tabcolsep)*140/1000\relax}>{\raggedright\arraybackslash}p{\dimexpr(\columnwidth-6\tabcolsep)*593/1000\relax}}\toprule
Component & Where & Role \\
\midrule
modelentry & CPU & Adapts protocols and keeps the client connection; forwards, queues or rejects as the dispatcher decides; owns the client-facing request identity; fronts the durable queue of the flex tier. \\
modeldispatcher & cluster & Holds the global view; makes every placement decision (forward, queue or reject a request) and drives engine scale-out and scale-in. \\
modellet & node & Hosts the engine processes and manages local GPU allocation; enforces that at most one engine is active on a card group; heals engines. \\
engine plugin & engine & Advertises which operations the engine supports; carries out sleep, wake and pause inside the engine process. \\
\bottomrule\end{tabular}
\end{table}

\begin{figure}[t]\centering\includegraphics[width=\columnwidth]{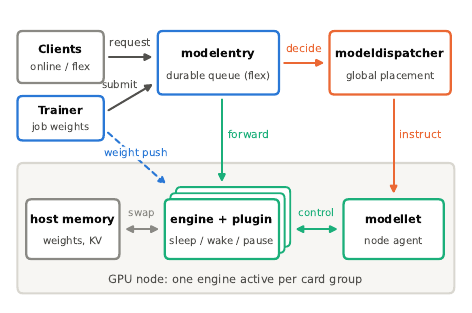}\caption{ModelLess architecture. One modellet per node manages several engines;
the dashed arrow is the trainer's weight push into job engines.}\end{figure}

\subsection{Requests}\label{requests}

A request reaches ModelLess in one of three forms, one for each tier of
§4.2, and each shows the control path end to end.

\textbf{Online.} A request arrives at modelentry, which registers it
with the dispatcher to ask for a serving engine. If an active engine is
below its concurrency threshold (§6.1), the dispatcher returns a forward
decision and the endpoint, and the request is forwarded at millisecond
cost. If all active engines have reached their target concurrency, the
dispatcher first tries to scale out: if a sleeping engine can wake in
seconds, it returns a queue decision, tells modellet to wake the engine,
and modelentry holds the connection until the engine reports healthy; if
scale-out requires a cold start or cannot proceed, the dispatcher
overflows requests across existing active engines up to their hard
concurrency limit before queueing or rejecting (§6.1). If the wait
exceeds its limit or no capacity can be obtained, the dispatcher rejects
the request, and the client may retry. The dispatcher separates the fast
per-request decision from the slower scaling of engines into two
independent loops joined by a task queue, so that neither blocks the
other.

\textbf{Job.} The training framework submits a task with a weight
version, a model, and \texttt{min}/\texttt{max}. The dispatcher chooses
card groups and starts the first \texttt{min} engines, marked as needing
a seal, with base weights, not the step's weights, because ModelLess
owns process, memory and scheduling while the training side owns pushing
each step's weights. Because the base runtime persists across training
steps, advancing to a new weight version rebinds runtimes on the same
base runtimes rather than restarting engine processes. The dispatcher
returns the engine endpoints as they start, and the training framework
forms its own communication group over these engines and pushes the
weights. The submission completes when the \texttt{min} engines have
received them, which ModelLess infers from the trainer's first
generation request to each engine; generation requests go through
modelentry, which forwards them, while the returned endpoints serve only
the trainer's weight push, so no extra protocol message is needed. After
the first successful generation with those weights, ModelLess seals the
engine's weights. When concurrency rises, the dispatcher adds engines up
to \texttt{max} by following a sealed engine, and the new engines can be
preempted by higher tiers. A release request puts the engines to sleep
and sends the weight version to garbage collection after its references
drop and a grace period passes. The number of weight versions kept per
model is bounded to protect host memory.

\textbf{Flex.} The client submits a request to the durable queue and
polls for its result. The dispatcher delivers it to an engine on cards
that are completely idle, since flex borrows only whole idle engines,
and modelentry records each attempt in the queue. If a higher tier needs
the cards, the attempt is fenced and the request is requeued, or paused
where the engine supports it (§5.4), and the client keeps polling the
same request.

\section{Instance State and Tier
Semantics}\label{instance-state-and-tier-semantics}

An LLM instance has four layers of state (§4.1), and what an eviction
costs depends on which of them survive. Each tier fixes which loss its
clients tolerate (§4.2). That determines what the platform keeps across
a preemption, which operation it uses (§5), and the order in which the
control plane acts (§6). The tier bounds the loss, and the operation
picks the cost within that bound. Table 4 follows one tier per column,
from what it tolerates to what the platform does. Job engines are
preempted after flex and only above \texttt{min}.

\begin{table*}[t]
\caption{Tiers and their eviction semantics.}\centering\footnotesize
\begin{tabular}{>{\raggedright\arraybackslash}p{\dimexpr(\textwidth-8\tabcolsep)*208/1000\relax}>{\raggedright\arraybackslash}p{\dimexpr(\textwidth-8\tabcolsep)*263/1000\relax}>{\raggedright\arraybackslash}p{\dimexpr(\textwidth-8\tabcolsep)*263/1000\relax}>{\raggedright\arraybackslash}p{\dimexpr(\textwidth-8\tabcolsep)*263/1000\relax}}\toprule
 & online & job & flex \\
\midrule
Delivery & synchronous, connection held & task submission & asynchronous, durable queue (submit/poll) \\
Tolerates & nothing in flight: never paused, migrated or preempted by a lower tier & slowdown and rework (migration), not an open-ended stall & stall and delay, not loss \\
State kept across a preemption & in flight: all (never preempted); parked after \ensuremath{\theta} idle: weights in host memory, not KV & weights and the KV cache of a slept engine; in-flight requests continue elsewhere from their generated tokens & weights and prefix cache (requeue), or also the in-flight requests (pause) \\
Operation & none in flight; sleep/wake when parked after \ensuremath{\theta} idle & recompute continuation, on engines above \texttt{min} only & fenced requeue, or pause where the engine supports it \\
Capacity declaration & the service's replica range; high is reserved & \texttt{min} (reserved) and \texttt{max} (expansion limit) & the service's replica range; uses what is left \\
Concurrency target & low \texttt{suggested} / \texttt{max} concurrency, protects latency (§6.1) & high \texttt{suggested} / \texttt{max} concurrency, protects throughput & high \texttt{suggested} / \texttt{max} concurrency, protects throughput \\
When capacity is preempted & never preempted by job or flex; \texttt{high} is placed before \texttt{default}, and a default engine that high needs is drained & requests on engines above \texttt{min} migrate (recompute continuation); \texttt{min} engines are never preempted & requests fenced and requeued; paused instead when the engine supports it \\
Typical use & user-facing serving & RL rollout, data generation & offline batch inference \\
Evaluation in §7 & trace replay (§7.1) + 1-node e2e (§7.3) & analytic only (Table 12); follow/seal not evaluated e2e & 1-node e2e (pause vs requeue, §7.3) + analytic (Table 12) \\
\bottomrule\end{tabular}
\end{table*}

\begin{table*}[t]
\caption{State layers kept across a preemption, by tier.}\centering\footnotesize
\begin{tabular}{>{\raggedright\arraybackslash}p{\dimexpr(\textwidth-10\tabcolsep)*230/1000\relax}>{\raggedright\arraybackslash}p{\dimexpr(\textwidth-10\tabcolsep)*103/1000\relax}>{\raggedright\arraybackslash}p{\dimexpr(\textwidth-10\tabcolsep)*206/1000\relax}>{\raggedright\arraybackslash}p{\dimexpr(\textwidth-10\tabcolsep)*230/1000\relax}>{\raggedright\arraybackslash}p{\dimexpr(\textwidth-10\tabcolsep)*230/1000\relax}}\toprule
Layer & online, in flight & online, parked after \ensuremath{\theta} idle & job & flex \\
\midrule
Execution (process, compiled artifacts) & kept & kept (sleep) & kept (sleep) & kept (sleep) \\
Weights & kept on GPU & kept in host memory & kept in host memory & kept in host memory \\
KV cache & kept & lost; the next request starts cold & kept in the host cache of the slept engine; migrated requests recompute it on the target & prefix cache kept by write-through (requeue); all KV kept (pause) \\
In-flight requests & never preempted & none in flight & continue elsewhere from their generated tokens, KV recomputed & requeued with progress lost, or paused and resumed \\
\bottomrule\end{tabular}
\end{table*}

\subsection{Instance States}\label{instance-states}

An inference instance carries four kinds of state, and every handover
mechanism is a decision about which of them to keep. The
\emph{execution} layer is the process and its compiled artifacts (CUDA
graphs, kernels, warmup). The \emph{weights} layer is the model
parameters, in GPU memory when the instance serves. The \emph{KV} layer
is the attention cache of requests being generated and of requests
recently finished. The \emph{request} layer is the in-flight requests
themselves: their token prefixes, sampling state, and the client
connections waiting on them.

Cold start rebuilds all four. A handover is cheap only if it keeps the
layers that are expensive to rebuild. Weights and execution state cost
minutes to reload and warm up (a GLM-5.2 model takes 244--277 s). KV and
request state cost little to hold but are lost the moment GPU memory is
released. The operations of §5 differ in which of these layers they
keep.

\subsection{Tiers for Requests}\label{tiers-for-requests}

A tier is more than a priority band: the three workloads also sit at
different latency--throughput operating points. Because LLM decode
latency degrades with batch size while throughput improves with it, an
online instance must stay at low concurrency to protect TTFT and
inter-token latency, whereas job and flex instances favor high
concurrency to saturate the GPU. Rather than running a separate
scheduler per product, ModelLess expresses this operating point through
each tier's \texttt{suggested} and \texttt{max} concurrency thresholds
(Table 4) and applies one unified dispatcher policy across all tiers
(§6.1). Each tier also fixes how work is delivered, what capacity is
declared, and what its work loses when its cards are preempted. The
implementation has three tiers with priority online \textgreater{} job
\textgreater{} flex, in the spirit of priority classes in cluster
managers \citep{verma2015borg}.

The tiers differ in which loss their clients can tolerate. A preemption
can cost a request its work (it is aborted), its progress (it is
recomputed), its speed (it continues, slower), or its time (it stalls
until capacity returns). \emph{Online} traffic in flight tolerates none
of these: a user is waiting on each token. So job and flex never pause,
migrate or preempt an online request. An online service that has been
idle for \ensuremath{\theta} seconds may be parked, and its first
request after that pays the recovery stall. \emph{Job} traffic is a
batch of requests issued on behalf of a trainer that waits for it, as in
RL rollout or data generation. A job cannot be paused, because the
trainer on the other side would be blocked. A blocked trainer idles its
own GPUs, which are often many times the size of the rollout model, so a
stall wastes a large multiple of the job's GPU-time. A job does tolerate
slowdown. Its requests on engines above \texttt{min} can be migrated.
They continue elsewhere after a bounded stall, the time to re-prefill,
instead of waiting for the same cards to return, and they run more
slowly. A job favors high concurrency over inter-token latency.
\emph{Flex} traffic has no persistent connection. Users submit to a
message queue and poll for progress, so a stall costs time and nothing
else. Only the loss of a request is unacceptable. Only flex clients can
wait out a stalled request, so pause belongs to flex.

\textbf{Online} traffic has the highest priority, and its eviction
semantics have two parts. While an online instance has traffic it is
never preempted by a lower tier. When it has received no request for
\ensuremath{\theta} seconds it may be parked: its weights stay in host
memory and its KV cache is dropped, and the first request afterwards
waits for the recovery latency C of §2.5. The trace study of §7.1
measures what this costs. Job and flex traffic never preempt an online
engine or its in-flight requests, and a lower tier does not share an
engine with online traffic. Flex borrows only engines that are
completely idle, and when online traffic arrives, flex gives back the
whole card group. Requests that arrive while an engine is waking wait in
the queue.

The tier has two levels. \emph{Online-high} is the reserved level: its
admission check guarantees logical capacity entitlement (the cluster
admits only as many high runtimes as can be active at the same time),
without requiring those runtimes to stay resident in GPU memory when
idle. Two high services that compete for the same cards could not both
be guaranteed to activate. High capacity serves online traffic only: job
engines are never placed on it, and flex occupants that borrow it are
paused or requeued at once when a high engine must activate.
\emph{Online-default} is the overcommitted level, stacked on cards that
a high engine may later need. A request to a default service can be
rejected when no capacity can be found, whereas a high service wakes
whenever the admission check made at its creation still holds. A high
engine never preempts a default engine; it drains it. The default engine
takes no new requests, finishes its in-flight requests and then hands
over its cards, so a high engine activates after the longest default
request plus the wake time C. Because long default generations delay
activation, stacking default services on high cards assumes bounded
default request durations or conservative overcommit; forcing an earlier
yield requires clients that tolerate migration (§8). An in-flight
default request is never migrated.

\textbf{Job} is for task-style throughput work. A submission declares
\texttt{min} and \texttt{max}. The \texttt{min} engines are a hard
reservation and an admission threshold: a submission does not complete
until at least \texttt{min} engines have received the job's weights. The
weight flow, from the push to the first generation, is described in
§3.4, and the sealing rule in §5.3. Engines above \texttt{min} are
expansion and are the ones preempted when a higher tier claims their
cards. \texttt{min} engines are never preempted, and before they are
ready they are also never put to sleep. A job engine keeps its KV cache
across a sleep, in the host cache tier of §5.4. An agentic workload
needs this: its requests share long prefixes across turns, and without
the cache every later turn on that engine would recompute its whole
context. Requests migrated to another engine do not benefit from it.
Because all active jobs' \texttt{min} values reduce the capacity
available to online bursts, \texttt{min} should be set to what the task
needs, not to as much as can be had.

\textbf{Flex} is for asynchronous work whose guarantee is that an
accepted request is not lost and completes once capacity is available,
with no bound on bandwidth or delay. Requests live in a durable queue,
so a preempted request is not lost (§5.4). Flex is the lowest tier: it
can be preempted by online and job traffic and preempts none.

Table 5 restates the semantics in the vocabulary of §4.1. It shows the
layers each tier forces the platform to keep, and the layers whose
retention is a cost choice, such as the request layer under pause or the
KV cache under write-through.

The platform's commercial products map onto the tiers: online products
belong to the online tier, asynchronous and batch products to flex, and
rollout to job. Online products with different latency targets differ in
how a service is configured, for example in how long it may idle before
it is put to sleep, and not in tier. ModelLess does not encode product
names.

Tier and weight source are orthogonal except in one case. Base weights
and derived weights loaded from a local path or a model hub may be
served in any tier. Weights pushed live by a trainer over a
collective-communication group are allowed only in the job tier, because
they depend on a training-side session (group creation, push,
validation) whose lifecycle only the job protocol provides. A job may
still use base weights: the job tier is a task protocol with reserved
and expandable capacity.

\section{Handover Operations}\label{handover-operations}

\subsection{Sleep and Wake}\label{sleep-and-wake}

\emph{Sleep} releases an instance's GPU memory while keeping its
execution layer and a copy of its weights in host memory, in the manner
of engine sleep modes and tensor-release libraries
\citep{vllm-sleep-mode, torch-memory-saver}. The instance stays in the
routing table, so a request that arrives while it sleeps triggers a wake
instead of a failure. \emph{Wake} restores the weights from host memory
into GPU memory; warmup and graph capture were done when the instance
was first prepared, so wake only copies. The preparation step (preheat)
runs each declared instance through one full cold start and then puts it
to sleep, which leaves a weight copy in host memory per instance; host
memory must be sized for this. On our 2 TB nodes, one GLM-5.2 borrower's
weight backup (373 GB) and configurable host KV pool plus the online
engine's backup occupy about 1.5 TB (§7.3.1); weight backups scale with
resident models, while the host KV pool is sized per deployment. Only a
request for a model that was never prepared, or for which every prepared
instance is busy, falls back to a cold start.

Sleep as defined here applies to an idle instance: modellet waits for
in-flight requests to finish and then releases. Calling the engine's own
generation pause for this purpose would freeze its scheduler and prevent
the drain from ever completing, so the plain path does not use it.
Measured cost is in §7.2. Draining waits for the longest in-flight
request, and aborting instead loses work, which tiers that must not lose
work cannot accept; the next operation exists for them.

\subsection{Pause and Resume}\label{pause-and-resume}

\emph{Pause} is sleep for an instance that still has requests in flight.
Instead of draining or discarding them, it keeps them. modellet closes
the instance's admission gate and then runs a checkpoint in a fixed
order. First, each in-flight request finishes its current step, and the
committed prefix is taken up to the last full page boundary. Second,
each request's exact token prefix, the committed and page-aligned part
of its KV cache (paged KV \citep{kwon2023vllm}), is written through to a
hierarchical cache in host memory
\citep{xie2025hicache, zheng2024sglang}. Third, those requests'
device-side KV is freed. Finally, the three GPU memory partitions
(weights, KV cache, graphs) are swapped out. \emph{Resume} runs the
reverse order: restore the partitions, restore each request's prefix
from host memory, rebuild the partial last page, rebuild the decode
batch, and restore sampling state such as the record of tokens seen for
repetition penalties. Each request continues where it stopped, without
re-prefilling its committed prefix. Only the tokens past the last full
page boundary need a rebuild, so the rebuild is under one page (64
tokens) per request in the common case; on GLM-5.2 the median was 33
tokens and the maximum 62 (§7.2). The engine's recomputed-token counter
counts committed tokens that the host backup does not cover, so it
excludes this rebuild, and it was zero in every microbenchmark cell
(§7.2).

\begin{table*}[t]
\caption{Operations and preemption paths by the state they keep. ``Cost'' refers
to measurements in §7.2.}\centering\footnotesize
\begin{tabular}{>{\raggedright\arraybackslash}p{\dimexpr(\textwidth-16\tabcolsep)*107/1000\relax}>{\raggedright\arraybackslash}p{\dimexpr(\textwidth-16\tabcolsep)*134/1000\relax}>{\raggedright\arraybackslash}p{\dimexpr(\textwidth-16\tabcolsep)*145/1000\relax}>{\raggedright\arraybackslash}p{\dimexpr(\textwidth-16\tabcolsep)*145/1000\relax}>{\raggedright\arraybackslash}p{\dimexpr(\textwidth-16\tabcolsep)*84/1000\relax}>{\raggedright\arraybackslash}p{\dimexpr(\textwidth-16\tabcolsep)*134/1000\relax}>{\raggedright\arraybackslash}p{\dimexpr(\textwidth-16\tabcolsep)*145/1000\relax}>{\raggedright\arraybackslash}p{\dimexpr(\textwidth-16\tabcolsep)*103/1000\relax}}\toprule
Mechanism & Applies to & Keeps & Discards & Requests in flight & Fails to & Cost & Eval. \\
\midrule
Sleep / wake & idle instance & execution, weights (host copy) & GPU memory; KV unless written through to the host cache (borrower engines, §5.4) & none (drained first) & rebuild or cold start & release \textasciitilde1 s, wake 2.9--4.4 s (GLM-5.2, TP8) & §7.2, §7.3 \\
Pause / resume & busy instance (capability required) & execution, weights, KV prefixes, requests, sampling counters & GPU memory, last partial page & kept, resumed exactly & recovery-required, admission closed & about 5--9 s per pause--resume, including KV reload from host (\textasciitilde3 \ensuremath{\mu}s/token) & §7.2, §7.3 \\
Requeue (flex) & async tier & weights; prefix cache if retained & decode progress (and prompt KV if cache not retained) & aborted, re-run & durable queue & lost decode, or full prompt recompute & §7.2, §7.3 \\
Recompute continuation (job) & above-min instances & weights & KV of migrated requests (a slept engine keeps its host KV) & re-prefilled elsewhere & bounded retries & re-prefill of prompt plus generated tokens & impl.; analytic only (§7.2) \\
Follow & scale-out & --- & --- & --- & refuse if source unsealed or stale & copy of sealed weights from a peer & impl.; not eval. e2e \\
\bottomrule\end{tabular}
\end{table*}

The checkpoint fails closed. The engine moves through \emph{active},
\emph{checkpointing}, \emph{sleeping} and \emph{restoring}; any step
that fails moves it to a \emph{recovery-required} state in which
admission stays closed until the instance is restored or rebuilt. An
instance never serves with a half-written checkpoint. A paused
instance's cards are reserved for its resume. The dispatcher does not
give them to another lower-tier task, and only the preempting tier may
use them. A checkpoint that waits for a resume longer than a fallback
period (fifteen minutes by default) is retired by fenced requeue (§5.4)
before it is discarded.

Pause is available only for engines that advertise the capability (the
engine plugin negotiates it by version and configuration). An engine
without it falls back to a plain sleep after a drain. Two semantic
limits apply. First, resume restores the exact prefix and the sampling
counters, so under deterministic inference a greedy request continues
token for token (verified on Qwen3-30B-A3B, and for single requests on
GLM-5.2, §7.2); with batch-dependent kernels it continues from the same
state up to numerical non-determinism. With temperature sampling, the
restored prefix and weights are bit-exact and the rebuilt partial page
is a small prefill, so each token is drawn from the same conditional
distribution up to numerical non-determinism, but the sampled path
differs from the uninterrupted one. §7.2 reports what we checked.
Second, a request in prefill is not checkpointed in the middle of a
prefill step: the pause waits for the running prefill step to finish, so
the added yield is bounded by one prefill step, which is one chunk when
chunked prefill is enabled.

\subsection{Follow and Seal}\label{follow-and-seal}

The weights an RL rollout engine serves have no home in storage: the
trainer produces a new version at every step and pushes it into the
engines over its own communication group
\citep{sheng2025hybridflow, fu2025areal, wang2025roll}, so the only
copies live in engines that have already received them. \emph{Follow}
lets a new engine start from such an engine instead of reading storage:
it copies the weights from a healthy instance of the same model and
weight version, one that has already received the step's weights. On one
node the source instance writes its weights as a read-only shared-memory
snapshot and the target maps it; across nodes the copy goes over RDMA.
\emph{Seal} is the rule that makes this safe against empty or
not-yet-pushed weights: only a weight version whose snapshot matches the
weights in GPU memory may act as a source, and the snapshot is taken
only after the instance has produced a successful generation with those
weights. An unsealed instance is never a source, because a copy from it
could carry empty or stale weights. The control interface is
asynchronous and idempotent per request identifier, rejects stale
instance identities, and fences both endpoints during the transfer.

\subsection{Preempting Capacity}\label{preempting-capacity}

When a higher tier needs cards that a lower tier holds, the fate of the
lower tier's in-flight requests depends on its tier, as summarized in
Table 6. Online traffic is never preempted by a lower tier, as in
priority schemes for production clusters \citep{verma2015borg}: no lower
tier preempts an online engine or its requests, and a lower tier borrows
only fully idle instances.

For the \emph{job} tier (rollout-style workloads), instances above the
declared minimum are preempted by \emph{recompute continuation}.
Resubmitting the prompt plus the tokens generated so far is close to the
token-level response migration of RLBoost \citep{wu2026rlboost}.
SpotServe's stateful recovery also resumes interrupted requests from
their committed tokens, but it migrates their KV cache between instances
\citep{miao2024spotserve}, as does Llumnix for running requests
\citep{sun2024llumnix}, while Mooncake moves KV between prefill and
decode pools \citep{qin2025mooncake}; our recompute continuation does
not move the KV cache. We recompute for two reasons: migrating KV adds
engineering complexity, and re-prefilling is fast enough for a job's
workload. The same limit applies to pause: its checkpoint lives in the
host memory of one node and resumes on the same card group, because
resuming elsewhere would be a KV migration. For recompute continuation,
modelentry accumulates the partial output a streaming request has
produced (prompt token identifiers, generated token identifiers, text)
and cancels the request. It then resubmits the prompt plus the generated
tokens to another instance of the same model and weight version, with
the generation budget reduced by what was already produced. The client
sees a complete output and a brief stall. A prefix cache
\citep{zheng2024sglang} that survives on the target reduces the cost,
and the loop is bounded (two attempts by default) to prevent migration
storms.

For the \emph{flex} tier (queued, asynchronous work), a preempted
request is \emph{fenced and requeued}: modelentry first marks the
current attempt in the durable queue with a compare-and-set and enqueues
a new attempt, and only then aborts the request on the engine. The order
guarantees that the request is neither lost (it has been re-enqueued)
nor delivered twice (the old attempt is already marked, so its result is
not accepted). Where the engine supports pause, a flex request can
instead be held in a checkpoint and resumed, which avoids both
repetition and recomputation; where it does not, requeue is the path.

Requeue and recompute continuation have conditions that pause does not.
Requeue is valid only before any output is delivered, so it fits
asynchronous work; a streaming request that restarted would repeat
output it already sent. Its cost also depends on the borrower's prefix
cache surviving the release: with the cache retained, a requeued prompt
is almost entirely cached and only decode progress is lost, and without
it the whole prompt is recomputed (§7.2). Only flex clients see no
output before a request completes, so requeue is restricted to flex.

The retained cache lives in the engine's hierarchical cache in host
memory \citep{xie2025hicache}, to which committed prefixes are written
through before a release. The community engine drops the prefix cache
when it sleeps, so a woken engine would re-prefill every prompt in full
(§7.2); the platform therefore writes committed prefixes through to the
host tier. With the write-through, a requeued prompt hits the host tier
and only the decode progress is lost; this makes requeue a stronger
baseline than a plain sleep would, and it is the same write-through that
pause relies on. An engine that dies loses the host tier with it, and
the replays that follow recompute their prompts in full.

\textbf{Preemption is safe only if the admission gate closes before
memory is released.} Between the decision to preempt and the completion
of the release, the engine must not accept new generations. A request
admitted into that interval would either pin state that the release
cannot account for, or be routed into an engine whose memory is already
gone. The pause path closes the gate as its first step; the requeue path
does the same explicitly. The protocol requires it.

\subsection{Summary}\label{summary}

Table 6 summarizes the operations and preemption paths by the state each
one keeps and discards, what happens to requests in flight, how each
fails, and what it costs.

\section{Control Plane and Safety}\label{control-plane-and-safety}

\subsection{Placement and Victim
Order}\label{placement-and-victim-order}

Placement in ModelLess is demand-driven: the dispatcher does not rely on
traffic forecasting to pre-warm capacity ahead of bursts, because
seconds-scale wake lets real request arrivals materialize sleeping
capacity directly. When a request needs an engine, the dispatcher routes
it using two per-tier concurrency thresholds---a soft \emph{suggested
concurrency} that marks an instance's latency--throughput operating
point (low for online, high for job and flex, Table 4) and a hard
\emph{max concurrency} ceiling---walking an ordered path:

\begin{enumerate}
\def\labelenumi{\arabic{enumi}.}
\tightlist
\item
  \textbf{Pack below suggested concurrency.} If one or more active
  engines are below their suggested concurrency, the dispatcher packs
  the request onto an active engine (preferring engines that already
  carry requests) so that spare engines stay idle, go to sleep, and can
  be lent to lower tiers.
\item
  \textbf{Scale out on reaching suggested concurrency.} Once every
  active engine reaches its suggested concurrency, the dispatcher
  attempts to add capacity before overloading existing engines: first by
  waking a sleeping engine of the same model and weight version
  (preempting any borrower that holds its cards), then by copying
  weights from a sealed active peer (follow, §5.3), and otherwise by
  preempting a lower-tier occupant (§5.4) and waking or cold-starting a
  new engine. When a fast wake is underway, the arriving request waits
  for the waking engine rather than pushing active engines past their
  suggested concurrency.
\item
  \textbf{Overflow round-robin on cold start or cap.} If scaling out
  requires a minutes-long cold start, or cannot proceed because the
  service has reached \texttt{max} replicas or no capacity can be
  reclaimed, the dispatcher does not block requests on the cold start;
  instead, it spills them across existing active engines in round-robin
  order up to each engine's hard max concurrency. Only when every active
  engine is at max concurrency are further requests queued or rejected.
\end{enumerate}

Online traffic is never preempted by a lower tier. Victims are taken in
tier order, flex occupants first and then job engines above
\texttt{min}, among online engines a default engine is drained for a
high one, and a job's \texttt{min} engines are never preempted.

\subsection{Safety Invariants}\label{safety-invariants}

Placement decides who gets the cards. The control plane keeps nine
invariants that make every change of ownership safe under concurrent
change. Each is enforced by the order of steps or by an identity check,
and each closes a specific race. We checked them by reading the step
order and identity checks, not by injecting faults.

\begin{enumerate}
\def\labelenumi{\arabic{enumi}.}
\tightlist
\item
  \textbf{Admission closes before memory is released.} The gate of an
  engine that is being preempted closes before any of its memory is
  freed (§5.4). A release first applies the same idle check as sleep. An
  engine can still be decoding the last tokens of a request whose client
  already has its answer, and a release that skipped the check would
  abort it. On wake, the gate stays closed until the engine has fully
  restored, so a wave of requeued requests cannot reach an engine that
  is still restoring. A release also checks that no device memory is
  left after the cache write-through.
\item
  \textbf{Fence before abort.} A flex request is marked in the durable
  queue with a compare-and-set and re-enqueued before its attempt on the
  engine is aborted, so it is neither lost nor delivered twice (§5.4). A
  drain or preemption takes the same compare-and-set fence on its
  source, so a state change since the decision aborts the action.
\item
  \textbf{Fail closed.} A failed pause step leaves the engine in a
  recovery-required state with admission closed, and an engine never
  serves from a half-written checkpoint (§5.2). A failed idle check also
  fails closed. A cache flush during a pause would invalidate the
  checkpoint, so pause and any operation that clears the cache are
  ordered against each other, and the checkpoint is a transaction.
\item
  \textbf{One active engine per card group.} modellet enforces this on
  every node.
\item
  \textbf{Reject stale instructions.} Every instruction names the exact
  runtime identity and incarnation, and modellet rejects any instruction
  whose incarnation is stale, because the engine has since been rebuilt.
\item
  \textbf{Never oversubscribe on wake.} Before waking an engine,
  modellet checks locally that the wake fits the node, and refuses it
  otherwise (§6.4).
\item
  \textbf{Seal before acting as a source.} An engine's weights may be
  copied only after it has produced a successful generation with them
  (§5.3).
\item
  \textbf{Quiet probes during transfer.} During a hardware-level
  transfer both endpoints are fenced and health probing is reduced to
  process liveness. A liveness probe that reads or writes memory that is
  moving can crash the engine.
\item
  \textbf{Online is never preempted by a lower tier, and a job's
  \texttt{min} engines are never preempted and, before they are ready,
  never put to sleep} (§4.2).
\end{enumerate}

\subsection{Engine States and
Failures}\label{engine-states-and-failures}

Three states are visible outside ModelLess: active, asleep, and
restoring. ModelLess also keeps \emph{creating}, \emph{failed} and
\emph{destroying}, each for a reason. Creating exists because a cold
start takes minutes, during which an engine can be neither routed to nor
started a second time. Failed exists so that a broken engine, which is
rebuilt, is distinguishable from a sleeping one, which is woken.
Destroying exists because a killed process takes time to exit, and an
engine that stays in it after its process has gone recovers without
restarting modellet. A failed pause step leaves the engine in the
recovery-required state of §5.2, and modellet restarts unhealthy engines
without operator action.

\subsection{Control-Plane Failures}\label{control-plane-failures}

Two rules keep a failure of the control plane from becoming a failure of
the data plane. First, before waking an engine, modellet checks locally
that the wake would not oversubscribe the node, and refuses the
instruction otherwise, so that even a mistaken or stale decision cannot
oversell the cards. Second, each modelentry keeps a list of the last
usable replicas, refreshed from the cluster metadata at an interval of
about 30 s. The dispatcher is a single instance; when it fails, every
modelentry enters a failover mode in which it forwards requests to the
replicas in its in-memory list, and all wakes and sleeps are disabled,
because no dispatcher is left to coordinate them and uncoordinated
changes would conflict. Serving continues on the engines that are
active, and capacity changes resume when the dispatcher returns. The
single dispatcher is therefore a point of failure for capacity changes:
during an outage a service that has scaled to zero cannot wake, so its
requests fail, and a borrower that holds a parked owner's cards cannot
be preempted, so that owner stays parked until the dispatcher returns.

\section{Evaluation}\label{evaluation}

We evaluate the online and flex tiers, and price the job tier
analytically (Table 12). §7.1 measures what lending costs the online
tier: how much capacity can be put to sleep, and what the first request
afterwards pays, as a function of the recovery latency C. §7.2 measures
what each layer of instance state costs to keep or rebuild. §7.3 runs
the flex tier on one node. It first checks that no flex request is lost
and every request completes, and then compares pause with requeue, two
ways of meeting the same tier semantics, to see what keeping the request
layer is worth.

\textbf{Trace study.} §7.1 replays the gateway request-arrival sequences
of the services described in §2.2 over two one-week periods, at
one-second resolution, with a discrete-event model of parking. It is a
simulation driven by production arrivals, not a deployment measurement.
\textbf{Mechanism measurements} (§7.2) run on 8 \ensuremath{\times} H20
(141 GB) nodes. Large-model results use GLM-5.2 (W4AFP8, 373 GB) with
tensor parallelism 8; smaller models use one to four cards. The pause
experiments use a frozen SGLang build with the engine-sleep extension.

\subsection{Trace Replay}\label{trace-replay}

\textbf{Method.} For each service and each idle threshold
\ensuremath{\theta} \ensuremath{\in} \{30, 60, 120, 300, 600, 1800\} s,
we replay the week continuously. After \ensuremath{\theta} seconds
without a request the service is parked; the next arrival triggers
recovery, which takes C seconds, and every request that arrives during
recovery also waits. We call the triggering request plus those waiting
arrivals the \emph{delayed} requests, and we judge every SLO gate on
delayed requests, because judging on triggers alone underestimates the
damage when C is large. We evaluate three recovery latencies. C = 5 s is
sleep/wake with weights kept in host memory. C = 60 s is an assumed
restart with weights preloaded in memory, which we have not measured. C
= 240 s is a restart that reads weights from disk; it is below our
measured cold starts of 244--277 s and so is conservative toward cold
start. Table 7 defines the gates. A service is \emph{reclaimable} at a
threshold if it passes the gate; its reclaimable share is its parked
share at the best passing threshold, and the fleet figure is the
card-weighted mean. Choosing the best threshold per service after the
fact is optimistic; we report fixed-\ensuremath{\theta} figures
alongside it, and compute 95\% confidence intervals by bootstrapping
over services.

\begin{table}[t]
\caption{SLO gates used in the replay. A gate is evaluated per service over the
delayed requests of the week. The P50, P90 and P99 gates bound the share
of requests that wait for recovery; they are not absolute TTFT
percentiles, which Figure 4 shows per service.}\centering\footnotesize
\begin{tabular}{>{\raggedright\arraybackslash}p{\dimexpr(\columnwidth-4\tabcolsep)*321/1000\relax}>{\raggedright\arraybackslash}p{\dimexpr(\columnwidth-4\tabcolsep)*678/1000\relax}}\toprule
Gate & Condition (per service) \\
\midrule
avg & mean TTFT increase \ensuremath{\leq} 2 s \\
P50 & delayed requests \textless{} 50\% of requests \\
P90 & delayed requests \ensuremath{\leq} 10\% \\
P99 & delayed requests \ensuremath{\leq} 1\% \\
timeout & 90 s timeout rate \ensuremath{\leq} 0.5\% (stricter SLO) or \ensuremath{\leq} 5\% (standard SLO) \\
median and timeout & P50 \ensuremath{\wedge} timeout (stricter SLO); P50 \ensuremath{\wedge} availability \ensuremath{\geq} 95\% (standard SLO) (availability taken as the complement of the 90 s timeout rate, so a timeout rate \ensuremath{\leq} 5\%) \\
\bottomrule\end{tabular}
\end{table}

\textbf{A cold start leaves little reclaimable capacity under an SLO.}
With C = 240 s, 69--85\% of services exceed the 0.5\% timeout threshold
(the range spans \ensuremath{\theta} from 300 s down to 30 s). Only
7.8\% {[}5.2, 10.9{]} of GPU-time is reclaimable under the
median-and-timeout gate (Table 8). With C = 240 s, a request that
arrives in the first 150 s of a recovery waits more than 90 s and times
out, so the 0.5\% clause caps the share of requests that arrive then.
Under the standard SLO (5\% timeouts, availability \ensuremath{\geq}
95\%) 16.4\% is reclaimable. A traditional restart is outside the SLO
for most services with traffic, and the services that pass hold little
GPU-time.

\textbf{Fast recovery moves the boundary, and how much depends on how
strict the SLO is.} Table 8 gives the reclaimable share per gate. Under
the median-and-timeout gate, C = 5 s and C = 60 s are close: 47.7\%
versus 42.5\%, a paired difference of 5.2 points {[}3.8, 6.8{]}. Under
experience-oriented gates the gap opens. The mean-TTFT gate shows 46.5\%
versus 18.1\% (a paired difference of 28.4 points {[}22.7, 33.5{]}). P90
shows 34.0\% versus 24.8\% (9.2 {[}5.3, 13.9{]}), and P99 shows 14.9\%
versus 11.2\% (3.6 {[}2.0, 5.6{]}). The coarse median-and-timeout gate
tolerates a minute of recovery, while a mean-TTFT gate needs seconds and
tail gates lose about a quarter of their capacity. The strictest tail
gate leaves little to reclaim at any C: at P99 only about 15\% of
GPU-time is reclaimable even at C = 5 s.

\begin{table*}[t]
\caption{Reclaimable share of GPU-time (\%), regular week, per-service best
\ensuremath{\theta}, with 95\% bootstrap intervals over services. C = 60
s is assumed, not measured. Paired differences in the text are computed
per service and need not equal the difference of the rounded shares.}\centering\footnotesize
\begin{tabular}{>{\raggedright\arraybackslash}p{\dimexpr(\textwidth-8\tabcolsep)*385/1000\relax}>{\raggedright\arraybackslash}p{\dimexpr(\textwidth-8\tabcolsep)*204/1000\relax}>{\raggedright\arraybackslash}p{\dimexpr(\textwidth-8\tabcolsep)*204/1000\relax}>{\raggedright\arraybackslash}p{\dimexpr(\textwidth-8\tabcolsep)*204/1000\relax}}\toprule
Gate \ensuremath{\uparrow} & C = 5 s & C = 60 s & C = 240 s \\
\midrule
avg \ensuremath{\leq} 2 s & 46.5 {[}41.5, 51.4{]} & 18.1 {[}14.5, 22.1{]} & 8.9 {[}6.2, 12.2{]} \\
P50 & 47.7 {[}42.6, 52.7{]} & 42.5 {[}38.1, 46.9{]} & 34.4 {[}30.0, 39.0{]} \\
P90 & 34.0 {[}29.9, 38.3{]} & 24.8 {[}20.3, 29.8{]} & 19.4 {[}15.4, 23.8{]} \\
P99 & 14.9 {[}11.3, 19.2{]} & 11.2 {[}8.4, 14.6{]} & 8.3 {[}5.5, 11.3{]} \\
timeout (90 s) & 52.6 {[}47.6, 57.6{]} & 49.9 {[}45.2, 55.2{]} & 7.8 {[}5.1, 10.9{]} \\
median and timeout, stricter SLO & 47.7 {[}43.1, 52.8{]} & 42.5 {[}37.5, 47.0{]} & 7.8 {[}5.2, 10.9{]} \\
median and timeout, standard SLO & 47.7 {[}42.8, 52.3{]} & 42.5 {[}37.7, 47.6{]} & 16.4 {[}13.0, 20.4{]} \\
\bottomrule\end{tabular}
\end{table*}

\textbf{The user-visible cost is concentrated.} At \ensuremath{\theta} =
120 s and C = 5 s, 0.10\% of requests are delayed and the mean TTFT
rises by 4 ms, so the fleet-level cost is negligible. The cost is not
spread evenly: for 10--20\% of services a majority of requests are
delayed (\ensuremath{\theta} from 120 s down to 30 s). Figure 4 shows it
per service, the price of scale-to-zero as a function of wake time. We
use engine-reported TTFT as the baseline because 99.8\% of gateway
requests are non-streaming, and the 2 s line applies to inputs under 4K
tokens.

\begin{figure*}[!htb]\centering\includegraphics[width=0.95\textwidth]{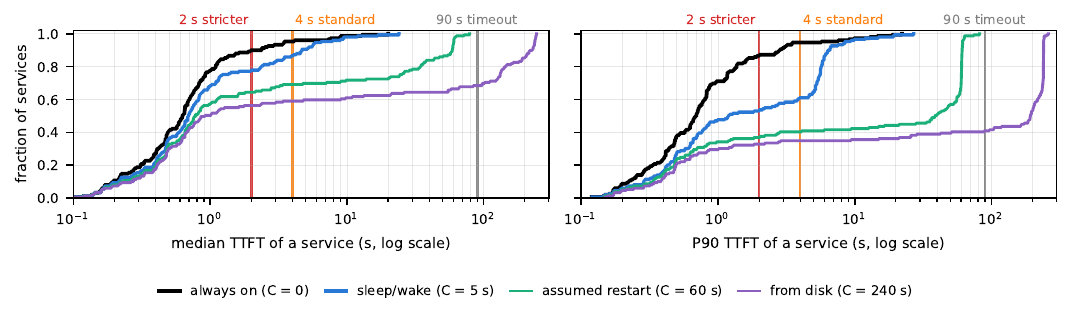}\caption{Per-service median and P90 TTFT under four recovery latencies against
the SLO lines (regular week, \ensuremath{\theta} = 60 s). The plot
covers the 149 of 212 services with TTFT data whose always-on median is
at least 100 ms.}\end{figure*}

With sleep/wake (C = 5 s, \ensuremath{\theta} = 60 s), 78\% of services
keep a median TTFT within 2 s, against 90\% when always on, and 87\%
stay within 4 s, against 95\%. At P90, 54\% stay within 2 s. With a cold
start (C = 240 s), 56\% stay within 2 s at the median, and a third
exceed the 90 s timeout.

\textbf{From gross to net.} Table 9 follows the funnel from gross parked
share to net reclaimable capacity. The gate costs far more than
overheads do. Applying the median-and-timeout gate keeps 91\% of the
ungated capacity with per-service thresholds and 81\% with
\ensuremath{\theta} fixed at 60 s. Subtracting borrower load and yield
time then removes only 0.6--3.4\% of what remains.

\begin{table*}[t]
\caption{From gross to net reclaimable capacity (regular week, 1,100 cards, C = 5
s, borrower load 5 s, yield 3 s, minimum useful segment 10 s).}\centering\footnotesize
\begin{tabular}{>{\raggedright\arraybackslash}p{\dimexpr(\textwidth-10\tabcolsep)*290/1000\relax}>{\raggedright\arraybackslash}p{\dimexpr(\textwidth-10\tabcolsep)*139/1000\relax}>{\raggedright\arraybackslash}p{\dimexpr(\textwidth-10\tabcolsep)*258/1000\relax}>{\raggedright\arraybackslash}p{\dimexpr(\textwidth-10\tabcolsep)*86/1000\relax}>{\raggedright\arraybackslash}p{\dimexpr(\textwidth-10\tabcolsep)*225/1000\relax}}\toprule
Threshold rule & Ungated gross & Median-and-timeout gated & Net & Net GPUs (equivalent) \\
\midrule
per-service best \ensuremath{\theta} & 52.6\% & 47.7\% & 46.1\% & 507 \\
fixed \ensuremath{\theta} = 60 s & 47.8\% & 38.8\% & 37.9\% & 417 \\
fixed \ensuremath{\theta} = 120 s & 42.9\% & 36.3\% & 35.9\% & 395 \\
fixed \ensuremath{\theta} = 300 s & 37.1\% & 33.6\% & 33.4\% & 368 \\
per-service best \ensuremath{\theta}, C = 60 s & --- & 42.5\% & 41.4\% & 455 \\
\bottomrule\end{tabular}
\end{table*}

The capacity is also steady. Figure 5 counts the cards that are lendable
in the same minute. The per-service best thresholds give a mean of 507
of the 1,100 cards and a 5th percentile of 342 (31\%).
\ensuremath{\theta} = 60 s gives a mean of 417 and a 5th percentile of
276. About 300 cards are lendable almost all of the time.

\begin{figure}[t]\centering\includegraphics[width=\columnwidth]{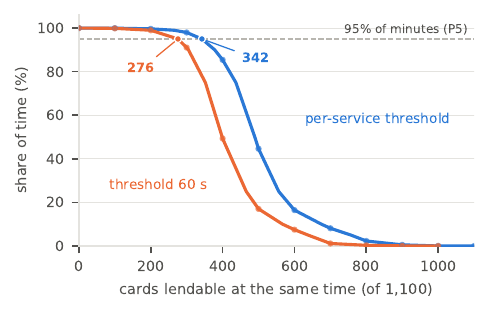}\caption{Share of the regular week in which at least X of the 1,100 cards are
lendable in the same minute.}\end{figure}

\textbf{Limits of this experiment.} The thresholds are chosen in-sample.
The replay parks a service after \ensuremath{\theta} seconds without a
new arrival, whereas the system waits for in-flight requests to finish,
so for services with long generations the replay overstates idle time.
The unit of lending is a whole service, so the result ignores node,
topology and fragmentation limits and cannot tell whether a service's
pods idle together. The borrower's yield time enters only the borrower's
side, and we did not rerun the owner's gates with C plus yield. The
release step measured in §7.3 has a median of 2.52 s in the pause run
and 1.56 s in the requeue run, consistent with the 3 s assumed here. C =
60 s is an assumption. Only gateway-visible traffic is observed.

\subsection{Mechanism Costs}\label{mechanism-costs}

\textbf{Sleep and wake.} Table 10 reports sleep (release) and wake
latencies. For GLM-5.2 with tensor parallelism 8 on H20, a steady-state
release takes 0.9--1.1 s and a wake 2.9--4.4 s, against a cold start of
244--277 s. A 32B-parameter model wakes in 0.4--0.7 s. The first release
after an engine start takes 80--108 s because host-memory backups of the
weights are built once; we report it separately and exclude it from
steady-state figures.

In a check on GLM-5.2 (tensor parallelism 8) with 100 GSM8K questions
(half greedy, half sampled at temperature 0.6), each question was run
after a full wake and followed by a sleep, 100 cycles in all. Accuracy
was 95\%, against 97\% and 98\% in two runs without sleeping, and the
greedy half scored 49/50 in every run, so the difference is within
run-to-run noise. Batch composition and cache state change the numerics,
so greedy token paths are not bit-reproducible across runs on this
engine, and we compare accuracy instead. Token identity after resume is
tested below. Decoding throughput was 71.6 against 71.7 tokens/s. Sleep
does drop the prefix cache, so each question paid a full re-prefill
after wake (about 0.25 s for an 800-token prompt). On long prompts that
cost matters, and pause's cache retention avoids it.

\begin{table}[t]
\caption{Handover latencies on H20 (steady state, seconds).}\centering\footnotesize
\begin{tabular}{>{\raggedright\arraybackslash}p{\dimexpr(\columnwidth-8\tabcolsep)*404/1000\relax}>{\raggedright\arraybackslash}p{\dimexpr(\columnwidth-8\tabcolsep)*191/1000\relax}>{\raggedright\arraybackslash}p{\dimexpr(\columnwidth-8\tabcolsep)*191/1000\relax}>{\raggedright\arraybackslash}p{\dimexpr(\columnwidth-8\tabcolsep)*212/1000\relax}}\toprule
Model (parallelism) & Release \ensuremath{\downarrow} & Wake \ensuremath{\downarrow} & Cold start \\
\midrule
GLM-5.2 (TP8) & 0.9--1.1 & 2.9--4.4 & 244--277 \\
32B-class (TP1--4) & --- & 0.4--0.7 & --- \\
\bottomrule\end{tabular}
\end{table}

\textbf{Pause and resume preserve in-flight requests.} Sleep must first
drain a busy engine, and releasing it without a drain, as requeue does,
aborts its running requests. Pause instead keeps each request's
committed prefix up to the last full page boundary, writes that KV cache
through to host memory, and releases weights and KV from the GPU. Resume
restores them, rebuilds the partial last page, and restores sampling
state. We measured this on GLM-5.2 (W4AFP8, TP8, 8 \ensuremath{\times}
H20-141 GB) with prompts of 4K--128K tokens and 1--8 concurrent
requests. Resume recomputed zero tokens in every cell (recomputed-token
count = 0). Under greedy decoding, the output after resume is identical
to the uninterrupted output for single requests in this microbenchmark.

We checked the correctness of pause on two models. On GLM-5.2 (TP8, 8
\ensuremath{\times} H20), 100 GSM8K questions (50 greedy, 50 sampled at
temperature 0.6) were each paused at their 30th, 60th or 100th generated
token and resumed after a full pause and resume cycle. Of the 100
answers, 99 were long enough to be paused. Accuracy was 98\% against
97\% for the same questions without pausing, and the greedy half scored
50/50 against 49/50, within run-to-run noise. Decoding throughput was
71.6 tokens/s, against 71.7 in the baseline of the sleep check. The
whole prefix cache survived: a median of 779 of about 800 prompt tokens
were cached after resume, whereas plain sleep and wake drop it. The
resumed requests rebuilt only their partial last pages, a median of 33
tokens and at most 62, under one 64-token page (3,407 tokens over the 99
resumes). The stream froze for the pause and resume transaction (median
6.9 s, at most 9.0 s) and then continued with no token lost or repeated.
A second check used a smaller model and a different configuration
(Qwen3-30B-A3B, TP = 2, page size 64, deterministic inference) on 100
GSM8K questions (50 greedy, 50 seeded samples at temperature 0.6, top-k
20). All 100 resumed sequences were identical to uninterrupted runs,
with bit-exact sampled log-probabilities and top-k entries after the
resume point. Accuracy was 90\% with and without pause, and five
requests resumed three times in succession were also exact.

Table 11 shows that the pause-plus-resume wall time is 5.05 s
(4.77--5.52, n = 5) for one 4K request and 7.25 s (6.66--7.75, n = 5)
for eight 32K requests, and temperature sampling gives the same costs
(5.03 s and 7.48 s). Reloading KV from host takes about 3
\ensuremath{\mu}s per token (12 ms for a 4K request, 878 ms for eight
32K requests, 445 ms for one 128K request). In contrast, recomputing a
context from scratch scales with its length: 0.48 s, 6.1 s and 61.7 s
for 4K, 32K and 128K tokens. Cold start of this engine takes four to
five minutes regardless of context.

\begin{table}[t]
\caption{Pause cost and the cost it avoids (GLM-5.2, W4AFP8, TP8). Pause + resume
is the wall time of the pause call plus the resume call and includes the
advance to a page boundary; the 4K \ensuremath{\times} 1 and 32K
\ensuremath{\times} 8 cells use the repeated protocol (n = 5), the other
cells the single-request protocol (n = 2).}\centering\footnotesize
\begin{tabular}{>{\raggedright\arraybackslash}p{\dimexpr(\columnwidth-8\tabcolsep)*113/1000\relax}>{\raggedright\arraybackslash}p{\dimexpr(\columnwidth-8\tabcolsep)*316/1000\relax}>{\raggedright\arraybackslash}p{\dimexpr(\columnwidth-8\tabcolsep)*240/1000\relax}>{\raggedright\arraybackslash}p{\dimexpr(\columnwidth-8\tabcolsep)*329/1000\relax}}\toprule
Setting & Pause + resume wall time \ensuremath{\downarrow} & KV reload from host & Recompute prefill instead \ensuremath{\downarrow} \\
\midrule
4K \ensuremath{\times} 1 & 5.05 s (4.77--5.52) & 12 ms & 0.48 s \\
32K \ensuremath{\times} 1 & 5.7 s & 107 ms & 6.1 s \\
32K \ensuremath{\times} 8 & 7.25 s (6.66--7.75) & 878 ms & 29.4 s mean (up to 50 s) \\
128K \ensuremath{\times} 1 & 6.0 s & 445 ms & 61.7 s \\
\bottomrule\end{tabular}
\end{table}

\emph{Reading the table.} Pause costs a few seconds that are dominated
by moving the weights, which sleep and wake pay too. What pause removes
is the context-dependent term, which exceeds a minute at 128K tokens.
This matters when the alternative is to recompute.

\textbf{What preempting a borrower costs depends on two properties of
the deployment.} When an owner returns, the borrower's in-flight
requests can be handled in one of three ways. \emph{Requeue} aborts the
request and runs it again from the start. \emph{Recompute continuation}
resubmits the prompt plus the tokens generated so far. \emph{Pause}
keeps the request. Their costs depend on whether the borrower's prefix
cache survives the release and on whether the request streams its
output. In our deployment the engine keeps the prefix cache across a
release, so a requeued request finds its prompt prefix cached (99.9\% of
the prompt, about 64 new tokens per replay in the engine logs) and loses
only its decode progress. Without this retention, a requeue recomputes
the prompt each time. In a one-hour requeue trial, 106 aborted requests
each replayed a cached prompt of about 65K tokens. Without retention we
estimate this at about 6.9M tokens of prefill, more than 2,000 s of
compute. Requeue is also valid only when no output has been delivered. A
streaming request has already flushed part of its output, and restarting
it would repeat that output. Pause provides both properties, since it
does not depend on the cache and keeps the request.

We quantify the compute lost to a preemption with an analytic model. It
draws 6,000 real lending windows (regular week, the same windows for
every scenario and concurrency; three draws) and runs a borrower
instance with B concurrent requests. Prefill is serialized at the
throughput we measured (9.4K, 5.6K and 2.3K tokens/s at 4K, 32K and
128K), and each policy is charged for the work it throws away. We use
the request lengths of the end-to-end workload of §7.3 (about 64K input,
4K output). Table 12 reports the lost share of usable GPU-time at the
two idle thresholds we use. With the cache retained, as deployed,
recompute continuation and pause lose 1--6\% and requeue loses 8--18\%.
Pause's advantage over recompute continuation is at most four points.
Without retention, requeue and recompute continuation lose 14--35\%,
more than ten times what pause loses. The mean window length varies by
about \ensuremath{\pm}20\% across draws, which moves the requeue losses
by a few points (standard deviation 2--7 points in the table's
scenarios). Differences of a point or two are within this noise. The
model is analytic: its request lengths are assumed, and the borrower's
decode rate is taken from a measurement of a different model.

\begin{table*}[t]
\caption{Share of usable GPU-time lost to preemption (\%, mean of three window
draws; analytic model; 64K-input / 4K-output requests; lower is better).}\centering\footnotesize
\begin{tabular}{>{\raggedright\arraybackslash}p{\dimexpr(\textwidth-12\tabcolsep)*240/1000\relax}>{\raggedright\arraybackslash}p{\dimexpr(\textwidth-12\tabcolsep)*101/1000\relax}>{\raggedright\arraybackslash}p{\dimexpr(\textwidth-12\tabcolsep)*164/1000\relax}>{\raggedright\arraybackslash}p{\dimexpr(\textwidth-12\tabcolsep)*113/1000\relax}>{\raggedright\arraybackslash}p{\dimexpr(\textwidth-12\tabcolsep)*278/1000\relax}>{\raggedright\arraybackslash}p{\dimexpr(\textwidth-12\tabcolsep)*101/1000\relax}}\toprule
Prefix cache & \ensuremath{\theta} (s) & Concurrency B & Requeue & Recompute continuation & Pause \\
\midrule
retained (deployed) & 30 & 8 & 12.5 & 2.5 & 1.5 \\
retained (deployed) & 30 & 32 & 18.4 & 6.0 & 2.0 \\
retained (deployed) & 60 & 8 & 8.2 & 1.5 & 0.9 \\
retained (deployed) & 60 & 32 & 12.5 & 3.7 & 1.2 \\
not retained & 30 & 8 & 23.0 & 19.7 & 1.5 \\
not retained & 30 & 32 & 34.6 & 33.0 & 2.0 \\
not retained & 60 & 8 & 16.3 & 13.8 & 0.9 \\
not retained & 60 & 32 & 26.7 & 25.4 & 1.2 \\
\bottomrule\end{tabular}
\end{table*}

\subsection{End-to-End Run}\label{end-to-end-run}

\subsubsection{Setup}\label{setup}

One node with eight H20 GPUs and 2 TB of host memory hosts two tenants
over the same cards. During the runs, the borrower's weight backup and
host KV pool and the online engine's backup occupy about 1.5 TB of host
memory. The online tenant is a stand-in for a production search-agent
service (Qwen3.5-35B-A3B, FP8, tensor parallelism 4, cards 4--7); we
replay its 336-request, 3.94 h arrival sequence through the gateway
path. The prompts are synthetic but reproduce the trace's request
lengths and prefix structure. The prefix cache is not forced to a target
hit rate and rebuilds as the engine runs. The borrower is GLM-5.2
(W4AFP8, tensor parallelism 8) occupying all eight cards whenever the
online engine is parked, fed by a durable queue with eight concurrent
requests of 64K input and 4K output tokens each (fixed seed). After 30 s
without an online request (\ensuremath{\theta} = 30 s) the online engine
is parked and its cards are lent to the borrower. We compare two ways of
handling the borrower's in-flight requests when an online request
arrives. \textbf{Requeue} aborts them and puts the requests back in the
queue. It is an abort-and-requeue baseline in the spirit of ROSE, which
aborts the rollout requests hit by a serving burst and reroutes them
\citep{gao2026rose}; it is not a reimplementation of ROSE, which
co-locates both models and reroutes aborted rollouts to other GPUs.
\textbf{Pause} (the pause run) checkpoints their KV cache to host memory
and resumes them where they stopped (§5.2, §5.4). The two runs use the
same binaries (the same build of every component), the same online
trace, the same borrower workload and the same seed. They differ in one
scheduler switch. The deployed requeue path keeps the prefix cache
across a release (§7.2). Each policy was run once for four hours,
requeue first and pause second, both starting from freshly started
engines with a cold cache.

\begin{table*}[t]
\caption{Four-hour controlled comparison (one run per policy; online tenant: 336
requests; borrower: 64K-in/4K-out requests, 8 concurrent).}\centering\footnotesize
\begin{tabular}{>{\raggedright\arraybackslash}p{\dimexpr(\textwidth-8\tabcolsep)*404/1000\relax}>{\raggedright\arraybackslash}p{\dimexpr(\textwidth-8\tabcolsep)*255/1000\relax}>{\raggedright\arraybackslash}p{\dimexpr(\textwidth-8\tabcolsep)*255/1000\relax}>{\raggedright\arraybackslash}p{\dimexpr(\textwidth-8\tabcolsep)*85/1000\relax}}\toprule
 & Pause & Requeue & Ratio \\
\midrule
Borrower output tokens \ensuremath{\uparrow} & 1,528,288 (375 requests) & 1,159,479 (285 requests) & 1.32 \\
Node-level GPU samples held by the borrower (nominally one per second) & 9,694 & 9,193 & 1.05 \\
Borrower tokens per held sample \ensuremath{\uparrow} & 157.7 & 126.1 & \textbf{1.25} \\
Preemptions of the borrower & 41 & 40 & --- \\
Samples with all eight GPUs free \ensuremath{\downarrow} & 0 & 198 & --- \\
Aborted request attempts (requeued) \ensuremath{\downarrow} & 0 & 310 & --- \\
Pool GPU samples busy (nvidia-smi util \textgreater{} 0) \ensuremath{\uparrow} & 81.6\% & 78.1\% & 1.04 \\
Online TTFT, median & 176 ms & 170 ms & \ensuremath{\approx} 1 \\
Online TTFT, 99th percentile & 10.1 s & 11.2 s & 0.90 \\
Online TTFT, maximum & 14.1 s & 82.7 s & --- \\
Engine failures (§7.3.2) \ensuremath{\downarrow} & 0 & 3 & --- \\
Online requests failed \ensuremath{\downarrow} & 0\% & 1.2\% (HTTP 504) & --- \\
\bottomrule\end{tabular}
\par\vspace{3pt}\begin{minipage}{\textwidth}\scriptsize\raggedright \emph{Note.} Output tokens are those of the delivered outputs of all
requests, all of which completed in both runs (the 310 aborted attempts
of the requeue run were re-run to completion). The requeue run lost
three engines, two of them under load specific to requeue and one in a
path both policies use; its four timeouts and its maximum TTFT of 82.7 s
occurred in the rebuild windows that followed.\end{minipage}
\end{table*}

\subsubsection{Results}\label{results}

\textbf{Pausing delivers more borrower work from the same node-level GPU
time (Table 13).} Per held sample, the pause run produced 25\% more
output tokens than requeue (1.32\ensuremath{\times} in total, because it
also held the GPUs for slightly longer). The sampler ran at about 0.83
Hz in the pause run and 0.81 Hz in the requeue run, so samples are not
seconds. Weighting each sample by its interval gives 11,733 and 11,460
held node-level seconds and a ratio of 1.29. The mechanism is the one §5
describes: requeue aborted 310 request attempts and discarded their
decode progress, while pausing resumed the same requests and rebuilt
only their partial last pages. On the online side, every request
succeeded in the pause run, and the median latency was the same in both
runs (176 ms and 170 ms).

\textbf{The flex tier's semantics held.} All 375 borrower requests of
the pause run and all 285 of the requeue run completed. None failed and
none was still in flight when the run ended. The 310 aborted attempts of
the requeue run were re-run to completion. No pause waited for its
resume longer than 224 s in the four-hour pause run, so the 15-minute
fallback never fired.

\textbf{The online tail is the price of the handover.} The 99th
percentile of online TTFT was 10.1 s in the pause run and 11.2 s in the
requeue run. In both runs about 12\% of requests had a first-token
latency above 2 s. A request that arrives after the cards have been lent
waits for the borrower to leave, for the online engine to wake, and for
its own prefill. The median-and-timeout gate of §7.1 holds for this
trace in the pause run. The requeue run's HTTP 504s (1.2\% of requests)
would fail the stricter SLO's 0.5\% timeout clause, and the P90 and P99
gates would fail in both runs.

The requeue run lost three engines during the run: two wake-side crashes
when a wave of requeued requests reached an engine that was still
restoring, and one release-side failure when HiCache demotion left
device KV unfreed on a path both policies share (every pause also writes
the cache through to the host tier, §5.2). The requeue run's 198 samples
with all eight GPUs free (Table 13), four HTTP 504 timeouts, and 82.7 s
maximum online TTFT all occurred in the rebuild windows that followed
these failures, which directly motivate the wake-side admission gate and
release-side residual-memory check of §6.2. We do not count the
resulting dead time as a benefit of pausing, so we report the ratio both
ways. Excluding, in both runs, the held GPU samples in which no kernel
ran (2.6\% of the requeue run's, 0.2\% of the pause run's) gives 1.22
instead of 1.25, and we treat 1.22 as the more conservative estimate.

\textbf{Limits of the comparison.} Each policy was run once, so
run-to-run variation is not measured; an earlier four-hour pair on a
pre-fix build (discarded for a deployment defect) and a one-hour pair
pointed the same way. The workload (64K input, 4K output) is long
relative to the lending windows and therefore favorable to a mechanism
that preserves progress. The comparison does not include requeue without
prefix-cache retention or recompute continuation, a job-tier operation
(§5.4). A sample counts as held when the borrower's GPU memory exceeds a
threshold. Samples are taken nominally every second (0.81 Hz in the
requeue run, 0.83 Hz in the pause run), and the per-sample ratio is
valid because both runs use the same method; the time-weighted ratio is
given above. The online tenant is a single instance of one service. Busy
GPU samples are samples in which nvidia-smi reports non-zero utilization
on a GPU. They show whether a kernel ran near that moment, not how much
of the GPU's throughput was used. They come from the same sampling in
both runs (about 4 h at 0.81--0.83 Hz, 8 GPUs) and are not comparable
with the log-based shares of §7.3.3.

\subsubsection{Pool Time}\label{pool-time}

We classified each second of the controlled four-hour pause run (14,184
s) from the engine logs. The borrower held the GPUs for 11,486 s. Of
these, 7,156 s were logged computation and 3,642 s were short gaps of at
most 3 s between consecutive prefill/decode log lines during continuous
step execution (together 94\% of held time); release steps took 104 s,
warm-up after wake 6 s and waiting 578 s. The online engine held its
GPUs for 2,484 s, of which 1,917 s were waiting, 296 s logged
computation, 17 s bridged gaps, 209 s health-probe prefills and 45 s
release steps. Weighted by the cards each engine uses (eight for the
borrower, four for the online engine), the pool computed for 51.5\% of
the run, against 1.0\% for the online service alone (52.2\% and 1.8\%
with the health probes); these shares count logged computation only, not
the bridged gaps. By the logs, 214 s of the run fall at handover edges
where neither engine held the GPUs, whereas the memory sampler of Table
13 saw no sample with all eight GPUs free; the two use different
signals. The reference is the online engine running alone on a node it
uses half of, which is how we modeled the production services we
observe; it is not a statically partitioned pool. The shares come from
log lines and GPU memory sampled at about 0.83 Hz, so they carry a
resolution of a few seconds, and the sampler's held time of §7.3.2
(11,733 s) differs from the log-based figure by about 2\%. The one-hour
structure and the comparison with the analytic model are in the Appendix
(§11).

\section{Discussion and Future Work}\label{discussion-and-future-work}

\textbf{What we found.} Long-tail online services leave about half their
GPU-time idle, mostly in windows of an hour or longer, and a
seconds-scale wake makes it lendable under a median-and-timeout SLO.
GLM-5.2 at TP8 sleeps in about 1 s and wakes in about 4 s. Pausing a
borrower with 64K-token requests yields 22\% more tokens per held GPU
sample than requeue. We evaluate the online and flex tiers; the job tier
is priced analytically (Table 12), and recompute continuation is not
exercised in §7.3.

\textbf{Where pause matters.} Pause pays when a preempted request has a
lot of decode progress to lose: a long-context request makes recompute
expensive, and a borrower preempted often inside a short window spends
its time rebuilding. It applies only to the flex tier, whose clients can
wait out a stall (§4.2). Recompute continuation, the operation of the
job tier, follows RLBoost, which migrates partial responses at the token
level by resubmitting the prompt plus the tokens generated so far
\citep{wu2026rlboost}. ModelLess applies it only to engines above
\texttt{min}. Requeue, continuation and pause differ in how much of the
request layer survives a preemption. With the prefix cache retained,
continuation recovers most of the gap to pause, and the analytic model
bounds pause's remaining margin at about four points (Table 12). The
choice among them follows the product form: what a client can tolerate
and what the engine supports. Pause adds exact state and independence
from the cache. A surviving cache and short requests favor requeue. Our
long workload (64K input, 4K output) favors keeping progress, so the
measured 1.22--1.25 ratio (1.22 conservative) is the benefit at that
point.

\textbf{Bring-up lessons.} Building and running the single-node
prototype surfaced five recurring pitfalls that shaped the safety
invariants of §6.2. First, calling an engine's generation pause during a
plain sleep freezes its scheduler and deadlocks the drain (§5.1).
Second, writing committed prefixes through to the host cache before
release turned requeue into a much stronger baseline than community
sleep, which drops the radix cache (§5.4). Third, liveness probes that
touch GPU memory during release or transfer crash the engine, so probing
must be reduced to process liveness while memory moves (§6.2). Fourth,
the three engine failures in the requeue run showed that admission must
stay closed until wake completes and that release must verify no device
KV remains after demotion (§6.2, §7.3.2). Fifth, health checks waking
idle engines block reclamation on the roughly 3,000 no-traffic cards
(§2.2), and an awake four-card online service still blocks all eight
cards during sub-\ensuremath{\theta} inter-arrival gaps (§7.3.3, §11.1),
motivating packing.

\textbf{Limitations.} Each experiment states its own limits (§2.4, §7.1,
§7.2, §7.3.2). Four apply across them. All measurements come from one
accelerator generation (H20), and the end-to-end result from one node,
one online trace and one borrower workload, with each policy run once.
The capacity figures are upper bounds, and the owner-side gates omit the
borrower's yield, so the gates that are sensitive to slow requests are
optimistic. Host memory, about 1.5 TB of a 2 TB node in our runs, bounds
how many owners and borrowers a node can hold. We did not run fault
injection, multi-pod simultaneity or multi-node experiments, and the
control-plane failover of §6.4 has not been measured under long
dispatcher outages.

\textbf{Future work.} Follow/seal needs to be evaluated with a trainer
in the loop, and at cluster scale it could extend overcommit, since a
new engine can start from a running one without reading storage. Idle
thresholds and card placement should also adapt to arrival patterns. A
streaming flex mode that tolerates network interruptions, for agent
harnesses, would widen the set of clients that can accept a stall. Once
KV can move between engines, through a pooling layer such as Mooncake or
LMCache or through peer-to-peer transfer, a preempted request could
resume on another engine instead of being recomputed. Packing instances
onto card groups would refine the lending unit from the whole service to
the card group. Forcing a default engine to yield its cards before its
requests drain needs synchronous clients that can tolerate a migration.
Sharing a card group between engines is limited by hardware, since a GPU
cannot yet be partitioned as finely as a CPU can under cgroups, so
capacity is planned on whole card groups. Pause also extends to
multi-step generative models such as diffusion transformers. At a
denoising-step boundary the state a request must keep is compact and
well defined: the latent, the step index, the solver history and the
sampler's random state; under sequence or patch parallelism
\citep{li2024distrifusion} every rank must checkpoint its shard
together. Diffusion serving systems already preempt and reschedule
requests at step boundaries inside one engine
\citep{lu2025tetriserve, ye2026genserve}, and approximate caching reuses
intermediate states across requests \citep{agarwal2024nirvana}. Pausing
at a step boundary and then releasing the whole GPU, weights included,
to another model would bring diffusion requests into the flex tier.
Feature caches that reuse activations across steps
\citep{ma2024deepcache, zhao2025pab, liu2025teacache, lv2025fastercache}
would play the role that the prefix cache plays for LLMs: dropping them
is safe but costs extra full steps after resume, so whether to keep them
is the same cost choice as keeping the KV cache. Finally, the analytic
model should be calibrated with per-service windows, so that a
deployment can predict whether pause is worth enabling before running
it.

\section{Related Work}\label{related-work}

\subsection{Cluster Managers and Preemptible
Resources}\label{cluster-managers-and-preemptible-resources}

Borg runs production and batch work on shared machines and preempts
lower-priority tasks to free resources \citep{verma2015borg}, an
approach continued by Omega and Kubernetes, which descend from Borg
\citep{schwarzkopf2013omega, burns2016borg}, while Mesos shares
resources among frameworks \citep{hindman2011mesos}. Harvest VMs expose
unallocated capacity to opportunistic workloads under explicit SLO
terms, with eviction as the price \citep{ambati2020harvest}; Heracles
co-locates best-effort work with latency-critical services by isolating
shared resources \citep{lo2015heracles}; Resource Central and Autopilot
predict and size workloads
\citep{cortez2017resourcecentral, rzadca2020autopilot}. Apart from
resizing, as Harvest VMs do when they shrink and grow, these systems
treat a task as a unit that is either running or evicted. An LLM
instance is different: what an eviction costs depends on which of its
weights, execution state, KV cache and in-flight requests are kept
(§4.1). Our contribution to this line is to expose that choice as
per-tier eviction semantics, not a new priority scheme.

\subsection{GPU Schedulers for Deep
Learning}\label{gpu-schedulers-for-deep-learning}

Gandiva, AntMan and Singularity time-slice, scale elastically, or
preempt and migrate training jobs transparently
\citep{xiao2018gandiva, xiao2020antman, shukla2022singularity}. Lyra is
the closest in spirit: it loans idle inference-cluster GPUs to elastic
training jobs and reclaims them \citep{li2023lyra}. We lend at the
granularity of a model instance, with seconds-scale preemption and
per-tier request semantics, and our borrowers include batch and rollout
inference. Workload studies of production GPU clusters characterize
utilization and queueing \citep{weng2022mlaas} and the failures and
scheduling of LLM development workloads \citep{hu2024acme}. Training
jobs typically checkpoint at iteration boundaries and tolerate minutes
of interruption, and Singularity's transparency operates at the job
level; an online inference instance must be restored in seconds and may
hold requests that cannot be re-run. Our trace study (§7.1) measures how
that constraint limits reclaimable capacity for the long tail of online
services.

Within a GPU, several systems let latency-sensitive and best-effort work
share a device. REEF, GPreempt and XSched add preemption of running GPU
work at kernel or queue granularity, and so rely on changes to the
launch path or the runtime
\citep{han2022reef, fan2025gpreempt, shen2025xsched}. Salus, Paella,
Orion and Tally share a GPU between jobs or requests through
fine-grained switching, kernel-level scheduling and interference control
\citep{yu2019salus, ng2023paella, strati2024orion, zhao2025tally}. These
mechanisms interleave or preempt work on a shared GPU at kernel or
command-queue granularity. Ours hands a whole card group to one instance
at a time and does not modify kernels or the GPU runtime (§5).

\subsection{LLM Serving Systems and Control
Planes}\label{llm-serving-systems-and-control-planes}

Serving engines and schedulers optimize a single deployment: continuous
batching, paged attention, prefill and decode scheduling and
disaggregation, and KV-cache pooling
\citep{yu2022orca, kwon2023vllm, zheng2024sglang, agrawal2024sarathi, zhong2024distserve, patel2024splitwise, qin2025mooncake}.
Llumnix migrates running requests and their KV caches between instances
for load balancing \citep{sun2024llumnix}; our continuation after a
preemption recomputes the context instead of moving it. AlpaServe,
MuxServe and QLM multiplex models or manage queues for SLO attainment
\citep{li2023alpaserve, duan2024muxserve, patke2024qlm}; DynamoLLM tunes
cluster configuration for energy and SpotServe serves on preemptible
instances \citep{stojkovic2025dynamollm, miao2024spotserve}.
Kubernetes-native control planes such as AIBrix and llm-d, and general
queueing and autoscaling components (Kueue, Volcano, KubeRay, KEDA,
Knative), manage and scale deployments of one product (queueing in Kueue
and Volcano, autoscaling in KubeRay, KEDA and Knative)
\citep{shan2025aibrix, llmd, kueue, volcano, kuberay, keda, knative-scale-to-zero}.
ModelLess adds a step these control planes do not provide: handing a
replica's GPUs to a lower tier and taking them back, with defined
semantics for what happens to the work in between.

\subsection{Multi-Model Pooling, Cold Start and
Keep-Alive}\label{multi-model-pooling-cold-start-and-keep-alive}

ModelLess belongs to the line of serverless LLM systems, which cut cold
start and pool GPUs among models. ModelLess keeps scale-to-zero and
hands the GPUs it frees to a lower tier under per-tier eviction
semantics. Prism shares GPU memory among many models, including
low-volume ones, by ballooning memory across them as their demand shifts
\citep{yu2025prism}. Aegaeon pools GPUs for concurrent long-tail models
with token-level switching \citep{xiang2025aegaeon}. We hand over whole
card groups instead, because a large tensor-parallel model leaves no
room for a second resident model (§2.6). PipeSwitch and Clockwork are
the closest earlier mechanisms for sleep/wake: both move model weights
between host memory and the GPU, PipeSwitch with pipelined context
switching that also lets one GPU alternate between training and
inference, and Clockwork with predictable load and execution times
\citep{bai2020pipeswitch, gujarati2020clockwork}. They target smaller
models and neither carries the state of requests in flight, so what we
add is the production LLM-engine path, the tier semantics, and the pause
and preemption order built around it. ConServe co-serves offline
requests in the same engine as online traffic, preempting them
layer-wise and resuming them from incrementally checkpointed KV cache
\citep{qiao2024conserve}. ServerlessLLM shortens model loading with a
fast checkpoint format, BlitzScale scales live by multicasting weights
between instances, \ensuremath{\lambda}Scale adds execute-while-load
scaling, and WarmServe prewarms instances
\citep{fu2024serverlessllm, zhang2025blitzscale, yu2025lambdascale, lou2026warmserve}.
Our idle-threshold policy is a keep-alive policy, a topic studied in
depth for serverless functions, where fixed windows are known to be
inferior to adaptive ones
\citep{shahrad2020serverless, fuerst2021faascache, roy2022icebreaker, lee2024spes}.
Fast-start and snapshot work shortens the restore itself
\citep{ustiugov2021reap, du2020catalyzer, oakes2018sock}; Dirigent cuts
orchestration latency \citep{cvetkovic2024dirigent}. We use fixed
thresholds, one \ensuremath{\theta} for all services or the best
\ensuremath{\theta} per service chosen in hindsight, and treat
adaptation as future work. Fast release and restore of an engine's
memory is available from the engines and from memory-saver tooling
\citep{vllm-sleep-mode, torch-memory-saver}. The difference from this
group lies in what the freed GPUs are for: we study handing them to a
lower tier under an SLO on the owner's side, and measure the capacity
that remains once recovery latency and SLO strictness are accounted for.

\subsection{Sharing Serving GPUs with RL
Rollout}\label{sharing-serving-gpus-with-rl-rollout}

RL post-training frameworks schedule and overlap rollout, reward and
training stages across engines
\citep{sheng2025hybridflow, hu2024openrlhf, fu2025areal, wang2025roll}.
ROSE places rollout on GPUs that serve online traffic, co-locating the
two under a controller that protects serving SLOs, aborts affected
rollouts on bursts and reroutes them, and reports end-to-end results
\citep{gao2026rose}. RLBoost harvests preemptible GPUs, including spot
instances and spare production capacity, for rollout and migrates
partial responses on preemption \citep{wu2026rlboost}. Our setting
differs in its unit of sharing, which is an instance on a card group
with exactly one model resident at a time, whereas ROSE keeps both
models resident and shares memory and compute spatially under
serving-first admission. Keeping a second model's weights resident takes
device memory that the online engine would otherwise use as KV cache, so
a large-model borrower would sharply cut the context the online engine
can hold; handing over the cards avoids this at the price of a
seconds-scale wake. We also treat batch and asynchronous inference, not
only rollout, as borrowers. We do not compare end-to-end results with
these systems.

\subsection{Memory Management and
Checkpoint/Restore}\label{memory-management-and-checkpointrestore}

Dynamic GPU memory management, including vAttention for serving and
GMLake for training, reduces fragmentation by decoupling virtual from
physical memory \citep{prabhu2025vattention, guo2024gmlake}. OS-level
GPU checkpoint and restore, such as PhoenixOS, CUDA's checkpoint utility
and CRIU's GPU support, snapshot a whole process
\citep{wei2025phoenixos, nvidia-cuda-checkpoint, criu-gpu}.
Request-level KV swap and recompute preemption exist inside engines
\citep{kwon2023vllm} and in co-serving systems \citep{qiao2024conserve},
and InferCept analyzes when to discard, preserve or swap the KV state of
requests that are interrupted \citep{abhyankar2024infercept}; the cost
comparison of Table 12 is of the same kind. Our pause differs in
releasing the whole GPU, weights included, across a handover to another
model. It is also narrower than process snapshots and engine-aware. It
writes through only the committed, page-aligned prefix of each request
and restores sampling state. Beyond the weight swap that sleep also
pays, its cost grows with the committed tokens (about 3
\ensuremath{\mu}s per token), not with a whole process image (§7.2).

\subsection{Workload Characterization}\label{workload-characterization}

Public and production LLM workload studies describe burstiness,
request-length distributions and caching behavior
\citep{wang2025burstgpt, xiang2026servegen, nixon2026yearllm}. SageServe
characterizes production LLM workloads at a cloud provider to drive
forecast-aware autoscaling \citep{jaiswal2025sageserve}. We contribute a
week-level view of the long tail of online services behind one gateway,
with arrival sequences at one-second resolution, in aggregate form (§2).

\section{Conclusion}\label{conclusion}

Idle GPU capacity in a multi-tier inference platform is large,
concentrated in long windows, and usable under strict latency gates only
if a handover takes seconds; a coarse median-and-timeout gate tolerates
about a minute, but a cold start of minutes leaves only 8\%. ModelLess
is a serverless system for this setting: users declare model services,
idle services scale to zero in seconds, and the GPUs they free go to
lower tiers. It makes the model instance the unit of scheduling and
preemption and gives each tier eviction semantics over the instance's
state. Four components (modelentry, modeldispatcher, modellet and an
engine plugin) enforce them under a preemption protocol that closes
admission before memory is released and fails closed. Sleep/wake,
pause/resume and follow/seal are the operations it uses. On one node,
preserving a preempted borrower's progress by pausing it yielded 22\%
more borrower tokens per held GPU sample than abort-and-requeue under
our trace and workload, without changing median online latency. These
results come from one run per policy on one accelerator generation, the
capacity figures are upper bounds from an offline replay, and the job
tier and follow/seal are not evaluated end to end (§8). Packing
instances onto card groups and adapting idle thresholds to arrival
patterns are the next steps.

\section{Appendix: One Lending Hour}\label{appendix-one-lending-hour}

\begin{figure*}[!htb]\centering\includegraphics[width=0.95\textwidth]{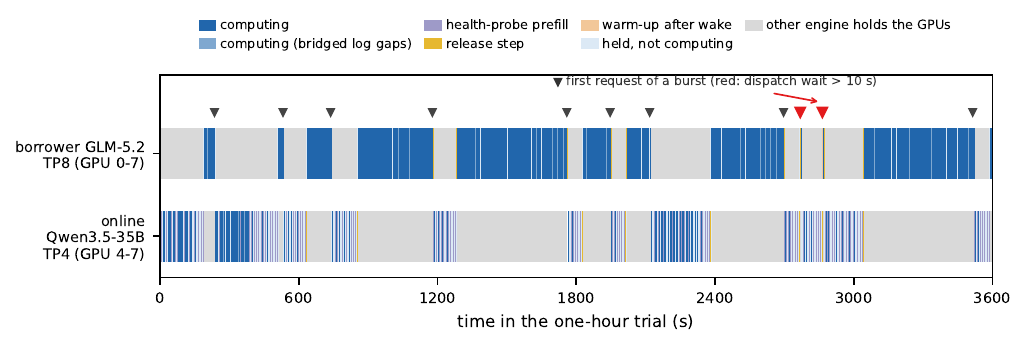}\caption{One hour of the pause policy (an hour at a higher request density than
the trace's average): the online engine (Qwen3.5-35B-A3B, GPUs 4--7) and
the borrower (GLM-5.2, all eight GPUs) alternate on the same GPUs.
Colors give the per-second state from the engine logs: computing (dark
blue, and lighter blue where a gap of at most 3 s between logged lines
is bridged), health-probe prefill, release step, warm-up after wake,
held but not computing, and gray where the other engine holds the GPUs.
Triangles mark the first request of a burst, which triggers a handover;
the two red ones had a dispatch wait above 10 s (15.3 s and 14.8 s).}\end{figure*}

\subsection{Timeline}\label{timeline}

Figure 6 shows one hour of the pause policy, second by second, from a
separate trial that is not part of the controlled comparison and ran at
a higher request density than the trace's average. The online engine and
the borrower alternate on the same GPUs over 23 handovers (12 lends and
11 preemptions of the borrower), and for only 1.6 s of the hour were the
GPUs held by neither engine. The borrower spent 93\% of the time it held
the GPUs computing, and the rest in release steps, warm-up after wake
and waiting. The online engine held its GPUs for 1,560 s but computed in
about a quarter of that time, because its requests are short and arrive
in clusters with gaps shorter than the idle threshold; this is why an
awake online engine leaves cards unused. Averaged over the pool, the
engines computed for 40.5\% of the hour against 5.1\% for the online
service alone.

\subsection{Model Against Measurement}\label{model-against-measurement}

The analytic model of §7.2, at the \ensuremath{\theta} and concurrency
of the four-hour run (\ensuremath{\theta} = 30 s, B = 8), has pause
ahead of requeue by about 1.12\ensuremath{\times}, against
1.22--1.25\ensuremath{\times} measured (§7.3.2). The direction and rough
size agree. Two explanations are plausible: this online service is
burstier than the fleet windows the model draws, so its preemptions
abort more work, and the engine deaths in the requeue run destroyed the
retained prefix cache, which the model prices at about
1.28\ensuremath{\times} when the cache is not retained. We did not test
either; the measured ratio holds for this trace and workload.

\clearpage

\bibliography{refs}

\begin{thebibliography}{86}
\providecommand{\natexlab}[1]{#1}
\providecommand{\url}[1]{\texttt{#1}}
\expandafter\ifx\csname urlstyle\endcsname\relax
  \providecommand{\doi}[1]{doi: #1}\else
  \providecommand{\doi}{doi: \begingroup \urlstyle{rm}\Url}\fi

\bibitem[Abhyankar et~al.(2024)Abhyankar, He, Srivatsa, Zhang, and
  Zhang]{abhyankar2024infercept}
R.~Abhyankar, Z.~He, V.~Srivatsa, H.~Zhang, and Y.~Zhang.
\newblock {InferCept}: Efficient intercept support for augmented large language
  model inference.
\newblock In \emph{Proceedings of the 41st International Conference on Machine
  Learning (ICML)}, 2024.
\newblock arXiv:2402.01869.

\bibitem[Agarwal et~al.(2024)Agarwal, Mitra, Chakraborty, Karanam, Mukherjee,
  and Saini]{agarwal2024nirvana}
S.~Agarwal, S.~Mitra, S.~Chakraborty, S.~Karanam, K.~Mukherjee, and S.~K.
  Saini.
\newblock Approximate caching for efficiently serving text-to-image diffusion
  models.
\newblock In \emph{21st USENIX Symposium on Networked Systems Design and
  Implementation (NSDI 24)}, 2024.
\newblock URL
  \url{https://www.usenix.org/conference/nsdi24/presentation/agarwal-shubham}.
\newblock arXiv:2312.04429.

\bibitem[Agrawal et~al.(2024)Agrawal, Kedia, Panwar,
  et~al.]{agrawal2024sarathi}
A.~Agrawal, N.~Kedia, A.~Panwar, et~al.
\newblock Taming throughput-latency tradeoff in {LLM} inference with
  {Sarathi-Serve}.
\newblock In \emph{18th USENIX Symposium on Operating Systems Design and
  Implementation (OSDI 24)}, pages 117--134, 2024.
\newblock URL
  \url{https://www.usenix.org/conference/osdi24/presentation/agrawal}.

\bibitem[Ambati et~al.(2020)Ambati, Goiri, Frujeri, et~al.]{ambati2020harvest}
P.~Ambati, {\'I}.~Goiri, F.~V. Frujeri, et~al.
\newblock Providing {SLOs} for resource-harvesting {VMs} in cloud platforms.
\newblock In \emph{14th USENIX Symposium on Operating Systems Design and
  Implementation (OSDI 20)}, pages 735--751, 2020.
\newblock URL
  \url{https://www.usenix.org/system/files/osdi20-paper-ambati.pdf}.

\bibitem[Bai et~al.(2020)Bai, Zhang, Zhu, and Jin]{bai2020pipeswitch}
Z.~Bai, Z.~Zhang, Y.~Zhu, and X.~Jin.
\newblock {PipeSwitch}: Fast pipelined context switching for deep learning
  applications.
\newblock In \emph{14th USENIX Symposium on Operating Systems Design and
  Implementation (OSDI 20)}, pages 499--514. USENIX Association, 2020.
\newblock URL \url{https://www.usenix.org/conference/osdi20/presentation/bai}.

\bibitem[Burns et~al.(2016)Burns, Grant, Oppenheimer, Brewer, and
  Wilkes]{burns2016borg}
B.~Burns, B.~Grant, D.~Oppenheimer, E.~Brewer, and J.~Wilkes.
\newblock Borg, omega, and {Kubernetes}.
\newblock \emph{ACM Queue}, 14\penalty0 (1):\penalty0 70--93, 2016.
\newblock \doi{10.1145/2898442.2898444}.
\newblock URL \url{https://research.google/pubs/borg-omega-and-kubernetes/}.

\bibitem[Cortez et~al.(2017)Cortez, Bonde, Muzio, Russinovich, Fontoura, and
  Bianchini]{cortez2017resourcecentral}
E.~Cortez, A.~Bonde, A.~Muzio, M.~Russinovich, M.~Fontoura, and R.~Bianchini.
\newblock Resource central: Understanding and predicting workloads for improved
  resource management in large cloud platforms.
\newblock In \emph{Proceedings of the 26th Symposium on Operating Systems
  Principles (SOSP '17)}, pages 153--167, 2017.
\newblock \doi{10.1145/3132747.3132772}.

\bibitem[{CRIU Project}(2026)]{criu-gpu}
{CRIU Project}.
\newblock {GPU} checkpointing -- {CRIU}.
\newblock \url{https://criu.org/GPU_Checkpointing}, 2026.
\newblock Accessed 2026-10-28.

\bibitem[Cvetkovi{\'c} et~al.(2024)Cvetkovi{\'c}, Costa, Djokic, Friedman, and
  Klimovic]{cvetkovic2024dirigent}
L.~Cvetkovi{\'c}, F.~Costa, M.~Djokic, M.~Friedman, and A.~Klimovic.
\newblock Dirigent: Lightweight serverless orchestration.
\newblock In \emph{Proceedings of the ACM SIGOPS 30th Symposium on Operating
  Systems Principles (SOSP '24)}, pages 369--384, 2024.
\newblock \doi{10.1145/3694715.3695966}.

\bibitem[Du et~al.(2020)Du, Yu, Xia, et~al.]{du2020catalyzer}
D.~Du, T.~Yu, Y.~Xia, et~al.
\newblock Catalyzer: Sub-millisecond startup for serverless computing with
  initialization-less booting.
\newblock In \emph{Proceedings of the Twenty-Fifth International Conference on
  Architectural Support for Programming Languages and Operating Systems (ASPLOS
  '20)}, pages 467--481, 2020.
\newblock \doi{10.1145/3373376.3378512}.

\bibitem[Duan et~al.(2024)Duan, Lu, Duanmu, et~al.]{duan2024muxserve}
J.~Duan, R.~Lu, H.~Duanmu, et~al.
\newblock {MuxServe}: Flexible spatial-temporal multiplexing for multiple {LLM}
  serving.
\newblock In \emph{Proceedings of the 41st International Conference on Machine
  Learning (ICML 2024)}, 2024.
\newblock URL \url{https://icml.cc/virtual/2024/poster/34064}.

\bibitem[Fan et~al.(2025)Fan, Ren, Xie, Gao, Shu, and Lu]{fan2025gpreempt}
R.~Fan, T.~Ren, M.~Xie, S.~Gao, J.~Shu, and Y.~Lu.
\newblock {GPreempt}: {GPU} preemptive scheduling made general and efficient.
\newblock In \emph{2025 USENIX Annual Technical Conference (USENIX ATC 25)},
  pages 263--272, 2025.
\newblock URL \url{https://www.usenix.org/conference/atc25/presentation/fan}.

\bibitem[Fu et~al.(2025)Fu, Gao, Shen, et~al.]{fu2025areal}
W.~Fu, J.~Gao, X.~Shen, et~al.
\newblock {AReaL}: A large-scale asynchronous reinforcement learning system for
  language reasoning.
\newblock In \emph{Advances in Neural Information Processing Systems 38
  (NeurIPS 2025)}, 2025.
\newblock URL \url{https://neurips.cc/virtual/2025/poster/117538}.

\bibitem[Fu et~al.(2024)Fu, Xue, Huang, et~al.]{fu2024serverlessllm}
Y.~Fu, L.~Xue, Y.~Huang, et~al.
\newblock {ServerlessLLM}: Low-latency serverless inference for large language
  models.
\newblock In \emph{18th USENIX Symposium on Operating Systems Design and
  Implementation (OSDI 24)}, pages 135--153, 2024.
\newblock URL \url{https://www.usenix.org/conference/osdi24/presentation/fu}.

\bibitem[Fuerst and Sharma(2021)]{fuerst2021faascache}
A.~Fuerst and P.~Sharma.
\newblock {FaasCache}: Keeping serverless computing alive with greedy-dual
  caching.
\newblock In \emph{Proceedings of the 26th ACM International Conference on
  Architectural Support for Programming Languages and Operating Systems (ASPLOS
  '21)}, pages 386--400, 2021.
\newblock \doi{10.1145/3445814.3446757}.

\bibitem[Gao et~al.(2026)Gao, Zhao, Muhtar, et~al.]{gao2026rose}
W.~Gao, Y.~Zhao, D.~Muhtar, et~al.
\newblock {ROSE}: Rollout on serving {GPUs} via cooperative elasticity for
  agentic {RL}, 2026.
\newblock URL \url{https://arxiv.org/abs/2605.06534}.

\bibitem[Gujarati et~al.(2020)Gujarati, Karimi, Alzayat,
  et~al.]{gujarati2020clockwork}
A.~Gujarati, R.~Karimi, S.~Alzayat, et~al.
\newblock Serving {DNNs} like {Clockwork}: Performance predictability from the
  bottom up.
\newblock In \emph{14th USENIX Symposium on Operating Systems Design and
  Implementation (OSDI 20)}, pages 443--462. USENIX Association, 2020.

\bibitem[Guo et~al.(2024)Guo, Zhang, Xu, et~al.]{guo2024gmlake}
C.~Guo, R.~Zhang, J.~Xu, et~al.
\newblock {GMLake}: Efficient and transparent {GPU} memory defragmentation for
  large-scale {DNN} training with virtual memory stitching.
\newblock In \emph{Proceedings of the 29th ACM International Conference on
  Architectural Support for Programming Languages and Operating Systems, Volume
  2 (ASPLOS '24)}, 2024.
\newblock \doi{10.1145/3620665.3640423}.

\bibitem[Han et~al.(2022)Han, Zhang, Chen, and Chen]{han2022reef}
M.~Han, H.~Zhang, R.~Chen, and H.~Chen.
\newblock Microsecond-scale preemption for concurrent {GPU}-accelerated {DNN}
  inferences.
\newblock In \emph{16th USENIX Symposium on Operating Systems Design and
  Implementation (OSDI 22)}, pages 539--558, 2022.
\newblock URL \url{https://www.usenix.org/conference/osdi22/presentation/han}.

\bibitem[Hindman et~al.(2011)Hindman, Konwinski, Zaharia,
  et~al.]{hindman2011mesos}
B.~Hindman, A.~Konwinski, M.~Zaharia, et~al.
\newblock Mesos: A platform for fine-grained resource sharing in the data
  center.
\newblock In \emph{8th USENIX Symposium on Networked Systems Design and
  Implementation (NSDI 11)}, pages 295--308, 2011.
\newblock URL
  \url{https://www.usenix.org/conference/nsdi11/mesos-platform-fine-grained-resource-sharing-data-center}.

\bibitem[Hu et~al.(2025)Hu, Wu, Shen, et~al.]{hu2024openrlhf}
J.~Hu, X.~Wu, W.~Shen, et~al.
\newblock {OpenRLHF}: A ray-based easy-to-use, scalable and high-performance
  {RLHF} framework.
\newblock In \emph{Proceedings of the 2025 Conference on Empirical Methods in
  Natural Language Processing: System Demonstrations}, pages 656--666, 2025.
\newblock \doi{10.18653/v1/2025.emnlp-demos.48}.
\newblock URL \url{https://arxiv.org/abs/2405.11143}.

\bibitem[Hu et~al.(2024)Hu, Ye, Wang, et~al.]{hu2024acme}
Q.~Hu, Z.~Ye, Z.~Wang, et~al.
\newblock Characterization of large language model development in the
  datacenter.
\newblock In \emph{21st USENIX Symposium on Networked Systems Design and
  Implementation (NSDI 24)}, pages 709--729, 2024.
\newblock URL \url{https://www.usenix.org/system/files/nsdi24-hu.pdf}.

\bibitem[Jaiswal et~al.(2025)Jaiswal, Jain, Simmhan,
  et~al.]{jaiswal2025sageserve}
S.~Jaiswal, K.~Jain, Y.~Simmhan, et~al.
\newblock {SageServe}: Optimizing {LLM} serving on cloud data centers with
  forecast aware auto-scaling.
\newblock \emph{Proceedings of the ACM on Measurement and Analysis of Computing
  Systems (POMACS)}, 9\penalty0 (3), 2025.
\newblock URL \url{https://arxiv.org/abs/2502.14617}.
\newblock arXiv:2502.14617; v1 titled ``Serving Models, Fast and Slow''.

\bibitem[Kabakibo et~al.(2026)Kabakibo, Trivedi, and
  Wang]{kabakibo2026breakingice}
H.~S. Kabakibo, A.~Trivedi, and L.~Wang.
\newblock Breaking the ice: Analyzing cold start latency in {vLLM}.
\newblock In \emph{Proceedings of the 9th Conference on Machine Learning and
  Systems (MLSys 2026)}, 2026.
\newblock URL \url{https://arxiv.org/abs/2606.07362}.
\newblock arXiv:2606.07362.

\bibitem[{KEDA Authors}(2026)]{keda}
{KEDA Authors}.
\newblock {KEDA}: Kubernetes event-driven autoscaling.
\newblock \url{https://keda.sh/}, 2026.
\newblock CNCF graduated project. Accessed 2026-10-28.

\bibitem[{Knative Authors}(2026)]{knative-scale-to-zero}
{Knative Authors}.
\newblock Configuring scale to zero -- {Knative}.
\newblock \url{https://knative.dev/docs/serving/autoscaling/scale-to-zero/},
  2026.
\newblock Accessed 2026-10-28.

\bibitem[{Kubernetes SIGs}(2026)]{kueue}
{Kubernetes SIGs}.
\newblock Kueue: Kubernetes-native job queueing.
\newblock \url{https://kueue.sigs.k8s.io/docs/overview/}, 2026.
\newblock Kubernetes SIGs project. Accessed 2026-10-28.

\bibitem[Kwon et~al.(2023)Kwon, Li, Zhuang, et~al.]{kwon2023vllm}
W.~Kwon, Z.~Li, S.~Zhuang, et~al.
\newblock Efficient memory management for large language model serving with
  {PagedAttention}.
\newblock In \emph{Proceedings of the 29th Symposium on Operating Systems
  Principles (SOSP '23)}, pages 611--626, 2023.
\newblock URL \url{https://arxiv.org/abs/2309.06180}.

\bibitem[Lee et~al.(2024)Lee, Zhu, Yang, et~al.]{lee2024spes}
C.~Lee, Z.~Zhu, T.~Yang, et~al.
\newblock {SPES}: Towards optimizing performance-resource trade-off for
  serverless functions.
\newblock In \emph{2024 IEEE 40th International Conference on Data Engineering
  (ICDE)}, pages 165--178, 2024.
\newblock \doi{10.1109/ICDE60146.2024.00020}.
\newblock URL \url{https://arxiv.org/abs/2403.17574}.

\bibitem[Li et~al.(2023{\natexlab{a}})Li, Xu, Zhu, Liu, Guo, and
  Wang]{li2023lyra}
J.~Li, H.~Xu, Y.~Zhu, Z.~Liu, C.~Guo, and C.~Wang.
\newblock Lyra: Elastic scheduling for deep learning clusters.
\newblock In \emph{Proceedings of the Eighteenth European Conference on
  Computer Systems (EuroSys '23)}, pages 835--850, 2023{\natexlab{a}}.
\newblock \doi{10.1145/3552326.3587445}.
\newblock URL \url{https://henryhxu.github.io/share/jiamin-eurosys23.pdf}.

\bibitem[Li et~al.(2024)Li, Cai, Cao, et~al.]{li2024distrifusion}
M.~Li, T.~Cai, J.~Cao, et~al.
\newblock {DistriFusion}: Distributed parallel inference for high-resolution
  diffusion models.
\newblock In \emph{IEEE/CVF Conference on Computer Vision and Pattern
  Recognition (CVPR)}, 2024.
\newblock URL \url{https://arxiv.org/abs/2402.19481}.
\newblock arXiv:2402.19481.

\bibitem[Li et~al.(2023{\natexlab{b}})Li, Zheng, Zhong,
  et~al.]{li2023alpaserve}
Z.~Li, L.~Zheng, Y.~Zhong, et~al.
\newblock {AlpaServe}: Statistical multiplexing with model parallelism for deep
  learning serving.
\newblock In \emph{17th USENIX Symposium on Operating Systems Design and
  Implementation (OSDI 23)}, pages 663--679, 2023{\natexlab{b}}.
\newblock URL
  \url{https://www.usenix.org/conference/osdi23/presentation/li-zhouhan}.

\bibitem[Liu et~al.(2025)Liu, Zhang, Wang, et~al.]{liu2025teacache}
F.~Liu, S.~Zhang, X.~Wang, et~al.
\newblock Timestep embedding tells: It's time to cache for video diffusion
  model.
\newblock In \emph{IEEE/CVF Conference on Computer Vision and Pattern
  Recognition (CVPR)}, 2025.
\newblock URL \url{https://arxiv.org/abs/2411.19108}.
\newblock arXiv:2411.19108.

\bibitem[{llm-d Project}(2026)]{llmd}
{llm-d Project}.
\newblock llm-d: Architecture.
\newblock \url{https://llm-d.ai/docs/architecture}, 2026.
\newblock CNCF Sandbox project. Accessed 2026-10-28.

\bibitem[Lo et~al.(2015)Lo, Cheng, Govindaraju, Ranganathan, and
  Kozyrakis]{lo2015heracles}
D.~Lo, L.~Cheng, R.~Govindaraju, P.~Ranganathan, and C.~Kozyrakis.
\newblock Heracles: Improving resource efficiency at scale.
\newblock In \emph{Proceedings of the 42nd Annual International Symposium on
  Computer Architecture (ISCA '15)}, pages 450--462, 2015.
\newblock \doi{10.1145/2749469.2749475}.

\bibitem[Lou et~al.(2026)Lou, Qi, Kang, et~al.]{lou2026warmserve}
C.~Lou, S.~Qi, R.~Kang, et~al.
\newblock {WarmServe}: Enabling one-for-many {GPU} prewarming for multi-{LLM}
  serving.
\newblock In \emph{International Conference on Machine Learning (ICML 2026)},
  2026.
\newblock URL \url{https://arxiv.org/abs/2512.09472}.
\newblock arXiv:2512.09472.

\bibitem[Lu et~al.(2025)Lu, He, Tan, et~al.]{lu2025tetriserve}
R.~Lu, S.~He, W.~Tan, et~al.
\newblock {TetriServe}: Efficiently serving mixed {DiT} workloads, 2025.
\newblock URL \url{https://arxiv.org/abs/2510.01565}.

\bibitem[Lv et~al.(2025)Lv, Si, Song, et~al.]{lv2025fastercache}
Z.~Lv, C.~Si, J.~Song, et~al.
\newblock {FasterCache}: Training-free video diffusion model acceleration with
  high quality.
\newblock In \emph{Thirteenth International Conference on Learning
  Representations (ICLR 2025)}, 2025.
\newblock URL \url{https://arxiv.org/abs/2410.19355}.
\newblock arXiv:2410.19355.

\bibitem[Ma et~al.(2024)Ma, Fang, and Wang]{ma2024deepcache}
X.~Ma, G.~Fang, and X.~Wang.
\newblock {DeepCache}: Accelerating diffusion models for free.
\newblock In \emph{IEEE/CVF Conference on Computer Vision and Pattern
  Recognition (CVPR)}, 2024.
\newblock URL \url{https://arxiv.org/abs/2312.00858}.
\newblock arXiv:2312.00858.

\bibitem[Miao et~al.(2024)Miao, Shi, Duan, et~al.]{miao2024spotserve}
X.~Miao, C.~Shi, J.~Duan, et~al.
\newblock {SpotServe}: Serving generative large language models on preemptible
  instances.
\newblock In \emph{Proceedings of the 29th ACM International Conference on
  Architectural Support for Programming Languages and Operating Systems, Volume
  2 (ASPLOS '24)}, pages 1112--1127, 2024.
\newblock \doi{10.1145/3620665.3640411}.

\bibitem[Ng et~al.(2023)Ng, Demoulin, and Liu]{ng2023paella}
K.~K.~W. Ng, H.~M. Demoulin, and V.~Liu.
\newblock Paella: Low-latency model serving with software-defined {GPU}
  scheduling.
\newblock In \emph{Proceedings of the 29th Symposium on Operating Systems
  Principles (SOSP '23)}, pages 595--610, 2023.
\newblock \doi{10.1145/3600006.3613163}.

\bibitem[Nixon et~al.(2026)Nixon, Durbin, Standhartinger, Gunawi, and
  Yang]{nixon2026yearllm}
W.~Nixon, J.~Durbin, F.~Standhartinger, H.~S. Gunawi, and J.~Yang.
\newblock A year in {LLM} serving: Workload evolution, caching and
  load-balancing, 2026.
\newblock URL \url{https://arxiv.org/abs/2608.13573}.

\bibitem[{NVIDIA}(2026)]{nvidia-cuda-checkpoint}
{NVIDIA}.
\newblock cuda-checkpoint: {CUDA} checkpoint and restore utility.
\newblock \url{https://github.com/NVIDIA/cuda-checkpoint}, 2026.
\newblock NVIDIA. Accessed 2026-10-28.

\bibitem[Oakes et~al.(2018)Oakes, Yang, Zhou, et~al.]{oakes2018sock}
E.~Oakes, L.~Yang, D.~Zhou, et~al.
\newblock {SOCK}: Rapid task provisioning with serverless-optimized containers.
\newblock In \emph{2018 USENIX Annual Technical Conference (USENIX ATC 18)},
  pages 57--70, 2018.
\newblock URL \url{https://www.usenix.org/conference/atc18/presentation/oakes}.

\bibitem[Patel et~al.(2024)Patel, Choukse, Zhang, et~al.]{patel2024splitwise}
P.~Patel, E.~Choukse, C.~Zhang, et~al.
\newblock Splitwise: Efficient generative {LLM} inference using phase
  splitting.
\newblock In \emph{2024 ACM/IEEE 51st Annual International Symposium on
  Computer Architecture (ISCA)}, pages 118--132, 2024.
\newblock \doi{10.1109/ISCA59077.2024.00019}.

\bibitem[Patke et~al.(2024)Patke, Reddy, Jha, et~al.]{patke2024qlm}
A.~Patke, D.~Reddy, S.~Jha, et~al.
\newblock Queue management for {SLO}-oriented large language model serving.
\newblock In \emph{Proceedings of the 2024 ACM Symposium on Cloud Computing
  (SoCC '24)}, pages 18--35, 2024.
\newblock \doi{10.1145/3698038.3698523}.
\newblock URL \url{https://arxiv.org/abs/2407.00047}.

\bibitem[Prabhu et~al.(2025)Prabhu, Nayak, Mohan, Ramjee, and
  Panwar]{prabhu2025vattention}
R.~Prabhu, A.~Nayak, J.~Mohan, R.~Ramjee, and A.~Panwar.
\newblock {vAttention}: Dynamic memory management for serving {LLMs} without
  {PagedAttention}.
\newblock In \emph{Proceedings of the 30th ACM International Conference on
  Architectural Support for Programming Languages and Operating Systems, Volume
  1 (ASPLOS '25)}, pages 1133--1150, 2025.
\newblock \doi{10.1145/3669940.3707256}.

\bibitem[Qiao et~al.(2026)Qiao, Anzai, Yu, et~al.]{qiao2024conserve}
Y.~Qiao, S.~Anzai, S.~Yu, et~al.
\newblock {ConServe}: Fine-grained {GPU} harvesting for {LLM} online and
  offline co-serving.
\newblock In \emph{Proceedings of the 43rd International Conference on Machine
  Learning (ICML 2026)}, 2026.
\newblock URL \url{https://arxiv.org/abs/2410.01228}.
\newblock arXiv:2410.01228.

\bibitem[Qin et~al.(2025)Qin, Li, He, et~al.]{qin2025mooncake}
R.~Qin, Z.~Li, W.~He, et~al.
\newblock Mooncake: Trading more storage for less computation -- a
  {KVCache}-centric architecture for serving {LLM} chatbot.
\newblock In \emph{23rd USENIX Conference on File and Storage Technologies
  (FAST 25)}, pages 155--170, 2025.
\newblock URL \url{https://www.usenix.org/system/files/fast25-qin.pdf}.

\bibitem[{Ray Project}(2026)]{kuberay}
{Ray Project}.
\newblock {Ray on Kubernetes (KubeRay)}.
\newblock \url{https://docs.ray.io/en/latest/cluster/kubernetes/index.html},
  2026.
\newblock Accessed 2026-10-28.

\bibitem[Roy et~al.(2022)Roy, Patel, and Tiwari]{roy2022icebreaker}
R.~B. Roy, T.~Patel, and D.~Tiwari.
\newblock {IceBreaker}: Warming serverless functions better with heterogeneity.
\newblock In \emph{Proceedings of the 27th ACM International Conference on
  Architectural Support for Programming Languages and Operating Systems (ASPLOS
  '22)}, pages 753--767, 2022.
\newblock \doi{10.1145/3503222.3507750}.

\bibitem[Rzadca et~al.(2020)Rzadca, Findeisen, Swiderski,
  et~al.]{rzadca2020autopilot}
K.~Rzadca, P.~Findeisen, J.~Swiderski, et~al.
\newblock Autopilot: Workload autoscaling at {Google}.
\newblock In \emph{Proceedings of the Fifteenth European Conference on Computer
  Systems (EuroSys '20)}, 2020.
\newblock \doi{10.1145/3342195.3387524}.

\bibitem[Schwarzkopf et~al.(2013)Schwarzkopf, Konwinski, Abd-El-Malek, and
  Wilkes]{schwarzkopf2013omega}
M.~Schwarzkopf, A.~Konwinski, M.~Abd-El-Malek, and J.~Wilkes.
\newblock Omega: Flexible, scalable schedulers for large compute clusters.
\newblock In \emph{Proceedings of the 8th ACM European Conference on Computer
  Systems (EuroSys '13)}, pages 351--364, 2013.
\newblock \doi{10.1145/2465351.2465386}.

\bibitem[Shahrad et~al.(2020)Shahrad, Fonseca, Goiri,
  et~al.]{shahrad2020serverless}
M.~Shahrad, R.~Fonseca, {\'I}.~Goiri, et~al.
\newblock Serverless in the wild: Characterizing and optimizing the serverless
  workload at a large cloud provider.
\newblock In \emph{2020 USENIX Annual Technical Conference (USENIX ATC 20)},
  pages 205--218, 2020.
\newblock URL
  \url{https://www.usenix.org/conference/atc20/presentation/shahrad}.

\bibitem[Shen et~al.(2025)Shen, Han, Liu, Chen, and Chen]{shen2025xsched}
W.~Shen, M.~Han, J.~Liu, R.~Chen, and H.~Chen.
\newblock {XSched}: Preemptive scheduling for diverse {XPUs}.
\newblock In \emph{19th USENIX Symposium on Operating Systems Design and
  Implementation (OSDI 25)}, pages 671--692, 2025.
\newblock URL
  \url{https://www.usenix.org/conference/osdi25/presentation/shen-weihang}.

\bibitem[Sheng et~al.(2025)Sheng, Zhang, Ye, et~al.]{sheng2025hybridflow}
G.~Sheng, C.~Zhang, Z.~Ye, et~al.
\newblock {HybridFlow}: A flexible and efficient {RLHF} framework.
\newblock In \emph{Proceedings of the Twentieth European Conference on Computer
  Systems (EuroSys '25)}, pages 1279--1297, 2025.
\newblock \doi{10.1145/3689031.3696075}.

\bibitem[Shukla et~al.(2022)Shukla, Sivathanu, Viswanatha,
  et~al.]{shukla2022singularity}
D.~Shukla, M.~Sivathanu, S.~Viswanatha, et~al.
\newblock Singularity: Planet-scale, preemptive and elastic scheduling of {AI}
  workloads, 2022.
\newblock URL \url{https://arxiv.org/abs/2202.07848}.

\bibitem[Stojkovic et~al.(2025)Stojkovic, Zhang, Goiri, Torrellas, and
  Choukse]{stojkovic2025dynamollm}
J.~Stojkovic, C.~Zhang, {\'I}.~Goiri, J.~Torrellas, and E.~Choukse.
\newblock {DynamoLLM}: Designing {LLM} inference clusters for performance and
  energy efficiency.
\newblock In \emph{2025 IEEE International Symposium on High Performance
  Computer Architecture (HPCA)}, pages 1348--1362, 2025.
\newblock \doi{10.1109/HPCA61900.2025.00102}.

\bibitem[Strati et~al.(2024)Strati, Ma, and Klimovic]{strati2024orion}
F.~Strati, X.~Ma, and A.~Klimovic.
\newblock Orion: Interference-aware, fine-grained {GPU} sharing for {ML}
  applications.
\newblock In \emph{Proceedings of the Nineteenth European Conference on
  Computer Systems (EuroSys '24)}, pages 1075--1092, 2024.
\newblock \doi{10.1145/3627703.3629578}.

\bibitem[Sun et~al.(2024)Sun, Huang, Zhao, et~al.]{sun2024llumnix}
B.~Sun, Z.~Huang, H.~Zhao, et~al.
\newblock Llumnix: Dynamic scheduling for large language model serving.
\newblock In \emph{18th USENIX Symposium on Operating Systems Design and
  Implementation (OSDI 24)}, pages 173--191, 2024.
\newblock URL
  \url{https://www.usenix.org/conference/osdi24/presentation/sun-biao}.

\bibitem[{The AIBrix Team} and Shan(2025)]{shan2025aibrix}
{The AIBrix Team} and J.~Shan.
\newblock {AIBrix}: Towards scalable, cost-effective large language model
  inference infrastructure, 2025.
\newblock URL \url{https://arxiv.org/abs/2504.03648}.

\bibitem[{torch\_memory\_saver contributors}(2026)]{torch-memory-saver}
{torch\_memory\_saver contributors}.
\newblock torch\_memory\_saver: allow tensor memory to be temporarily released
  and resumed later.
\newblock \url{https://github.com/fzyzcjy/torch_memory_saver}, 2026.
\newblock GitHub repository (MIT). Accessed 2026-10-28.

\bibitem[Ustiugov et~al.(2021)Ustiugov, Petrov, Kogias, Bugnion, and
  Grot]{ustiugov2021reap}
D.~Ustiugov, P.~Petrov, M.~Kogias, E.~Bugnion, and B.~Grot.
\newblock Benchmarking, analysis, and optimization of serverless function
  snapshots.
\newblock In \emph{Proceedings of the 26th ACM International Conference on
  Architectural Support for Programming Languages and Operating Systems (ASPLOS
  '21)}, pages 559--572, 2021.
\newblock \doi{10.1145/3445814.3446714}.

\bibitem[Verma et~al.(2015)Verma, Pedrosa, Korupolu, Oppenheimer, Tune, and
  Wilkes]{verma2015borg}
A.~Verma, L.~Pedrosa, M.~Korupolu, D.~Oppenheimer, E.~Tune, and J.~Wilkes.
\newblock Large-scale cluster management at {Google} with {Borg}.
\newblock In \emph{Proceedings of the Tenth European Conference on Computer
  Systems (EuroSys '15)}. ACM, 2015.
\newblock \doi{10.1145/2741948.2741964}.
\newblock URL
  \url{https://research.google/pubs/large-scale-cluster-management-at-google-with-borg/}.

\bibitem[{vLLM Project}(2026)]{vllm-sleep-mode}
{vLLM Project}.
\newblock Sleep mode -- {vLLM} documentation.
\newblock \url{https://docs.vllm.ai/en/latest/features/sleep_mode.html}, 2026.
\newblock Accessed 2026-10-28.

\bibitem[{Volcano Authors}(2026)]{volcano}
{Volcano Authors}.
\newblock Volcano: Cloud native batch scheduling system.
\newblock \url{https://volcano.sh/en/docs/}, 2026.
\newblock CNCF incubating project. Accessed 2026-10-28.

\bibitem[Wang et~al.(2025{\natexlab{a}})Wang, Xiong, and Chen]{wang2025roll}
W.~Wang, S.~Xiong, and G.~Chen.
\newblock Reinforcement learning optimization for large-scale learning: An
  efficient and user-friendly scaling library, 2025{\natexlab{a}}.
\newblock URL \url{https://arxiv.org/abs/2506.06122}.

\bibitem[Wang et~al.(2025{\natexlab{b}})Wang, Chen, Li,
  et~al.]{wang2025burstgpt}
Y.~Wang, Y.~Chen, Z.~Li, et~al.
\newblock {BurstGPT}: A real-world workload dataset to optimize {LLM} serving
  systems.
\newblock In \emph{Proceedings of the 31st ACM SIGKDD Conference on Knowledge
  Discovery and Data Mining V.2 (KDD '25)}, pages 5831--5841,
  2025{\natexlab{b}}.
\newblock \doi{10.1145/3711896.3737413}.

\bibitem[Wei et~al.(2025)Wei, Huang, Sun, et~al.]{wei2025phoenixos}
X.~Wei, Z.~Huang, T.~Sun, et~al.
\newblock {PhoenixOS}: Concurrent {OS}-level {GPU} checkpoint and restore with
  validated speculation.
\newblock In \emph{Proceedings of the ACM SIGOPS 31st Symposium on Operating
  Systems Principles (SOSP '25)}, pages 996--1013, 2025.
\newblock \doi{10.1145/3731569.3764813}.

\bibitem[Weng et~al.(2022)Weng, Xiao, Yu, et~al.]{weng2022mlaas}
Q.~Weng, W.~Xiao, Y.~Yu, et~al.
\newblock {MLaaS} in the wild: Workload analysis and scheduling in large-scale
  heterogeneous {GPU} clusters.
\newblock In \emph{19th USENIX Symposium on Networked Systems Design and
  Implementation (NSDI 22)}, pages 945--960, 2022.
\newblock URL \url{https://www.usenix.org/conference/nsdi22/presentation/weng}.

\bibitem[Wu et~al.(2026)Wu, Liu, Zheng, et~al.]{wu2026rlboost}
Y.~Wu, X.~Liu, H.~Zheng, et~al.
\newblock {RLBoost}: Harvesting preemptible cloud resources for cost-efficient
  reinforcement learning on {LLMs}.
\newblock In \emph{23rd USENIX Symposium on Networked Systems Design and
  Implementation (NSDI 26)}, pages 1323--1339, 2026.
\newblock URL \url{https://www.usenix.org/system/files/nsdi26-wu-yongji.pdf}.
\newblock arXiv:2510.19225.

\bibitem[Xiang et~al.(2025)Xiang, Li, Qian, et~al.]{xiang2025aegaeon}
Y.~Xiang, X.~Li, K.~Qian, et~al.
\newblock Aegaeon: Effective {GPU} pooling for concurrent {LLM} serving on the
  market.
\newblock In \emph{Proceedings of the ACM SIGOPS 31st Symposium on Operating
  Systems Principles (SOSP '25)}, pages 1030--1045, 2025.
\newblock \doi{10.1145/3731569.3764815}.
\newblock URL \url{https://ennanzhai.github.io/pub/sosp25-aegaeon.pdf}.

\bibitem[Xiang et~al.(2026)Xiang, Li, Qian, Yu, Zhai, and
  Jin]{xiang2026servegen}
Y.~Xiang, X.~Li, K.~Qian, W.~Yu, E.~Zhai, and X.~Jin.
\newblock {ServeGen}: Workload characterization and generation of large
  language model serving in production.
\newblock In \emph{23rd USENIX Symposium on Networked Systems Design and
  Implementation (NSDI 26)}, 2026.
\newblock URL \url{https://arxiv.org/abs/2505.09999}.
\newblock arXiv:2505.09999.

\bibitem[Xiao et~al.(2018)Xiao, Bhardwaj, Ramjee, et~al.]{xiao2018gandiva}
W.~Xiao, R.~Bhardwaj, R.~Ramjee, et~al.
\newblock Gandiva: Introspective cluster scheduling for deep learning.
\newblock In \emph{13th USENIX Symposium on Operating Systems Design and
  Implementation (OSDI 18)}, pages 595--610, 2018.
\newblock URL \url{https://www.usenix.org/conference/osdi18/presentation/xiao}.

\bibitem[Xiao et~al.(2020)Xiao, Ren, Li, et~al.]{xiao2020antman}
W.~Xiao, S.~Ren, Y.~Li, et~al.
\newblock {AntMan}: Dynamic scaling on {GPU} clusters for deep learning.
\newblock In \emph{14th USENIX Symposium on Operating Systems Design and
  Implementation (OSDI 20)}, pages 533--548, 2020.
\newblock URL \url{https://www.usenix.org/conference/osdi20/presentation/xiao}.

\bibitem[Xie(2025)]{xie2025hicache}
Z.~Xie.
\newblock {SGLang HiCache}: Fast hierarchical {KV} caching with your favorite
  storage backends.
\newblock \url{https://lmsys.org/blog/2025-09-10-sglang-hicache/}, 2025.
\newblock LMSYS Org blog, 2025-09-10. Accessed 2026-11-05.

\bibitem[Ye et~al.(2026)Ye, Li, Zhong, et~al.]{ye2026genserve}
F.~Ye, Z.~Li, X.~Zhong, et~al.
\newblock {GenServe}: Efficient co-serving of heterogeneous diffusion model
  workloads, 2026.
\newblock URL \url{https://arxiv.org/abs/2604.04335}.

\bibitem[Yu et~al.(2022)Yu, Jeong, Kim, Kim, and Chun]{yu2022orca}
G.-I. Yu, J.~S. Jeong, G.-W. Kim, S.~Kim, and B.-G. Chun.
\newblock Orca: A distributed serving system for transformer-based generative
  models.
\newblock In \emph{16th USENIX Symposium on Operating Systems Design and
  Implementation (OSDI 22)}, pages 521--538, 2022.
\newblock URL \url{https://www.usenix.org/conference/osdi22/presentation/yu}.

\bibitem[Yu et~al.(2025)Yu, Yang, Jia, et~al.]{yu2025lambdascale}
M.~Yu, R.~Yang, C.~Jia, et~al.
\newblock $\lambda${Scale}: Enabling fast scaling for serverless large language
  model inference, 2025.
\newblock URL \url{https://arxiv.org/abs/2502.09922}.

\bibitem[Yu and Chowdhury(2020)]{yu2019salus}
P.~Yu and M.~Chowdhury.
\newblock Salus: Fine-grained {GPU} sharing primitives for deep learning
  applications.
\newblock In \emph{Proceedings of Machine Learning and Systems (MLSys 2020)},
  2020.
\newblock URL \url{https://arxiv.org/abs/1902.04610}.
\newblock arXiv:1902.04610.

\bibitem[Yu et~al.(2026)Yu, Qiao, Ma, et~al.]{yu2025prism}
S.~Yu, Y.~Qiao, M.~Ma, et~al.
\newblock Prism: Cost-efficient multi-{LLM} serving via {GPU} memory
  ballooning.
\newblock In \emph{20th USENIX Symposium on Operating Systems Design and
  Implementation (OSDI 26)}, pages 75--93, 2026.
\newblock URL
  \url{https://www.usenix.org/conference/osdi26/presentation/yu-shan}.
\newblock arXiv:2505.04021.

\bibitem[Zhang et~al.(2025)Zhang, Wang, Liu, et~al.]{zhang2025blitzscale}
D.~Zhang, H.~Wang, Y.~Liu, et~al.
\newblock {BlitzScale}: Fast and live large model autoscaling with {O(1)} host
  caching.
\newblock In \emph{19th USENIX Symposium on Operating Systems Design and
  Implementation (OSDI 25)}, pages 275--293, 2025.
\newblock URL
  \url{https://www.usenix.org/conference/osdi25/presentation/zhang-dingyan}.

\bibitem[Zhao et~al.(2025{\natexlab{a}})Zhao, Jayarajan, and
  Pekhimenko]{zhao2025tally}
W.~Zhao, A.~Jayarajan, and G.~Pekhimenko.
\newblock Tally: Non-intrusive performance isolation for concurrent deep
  learning workloads.
\newblock In \emph{Proceedings of the 30th ACM International Conference on
  Architectural Support for Programming Languages and Operating Systems, Volume
  1 (ASPLOS '25)}, pages 1052--1068, 2025{\natexlab{a}}.
\newblock \doi{10.1145/3669940.3707282}.

\bibitem[Zhao et~al.(2025{\natexlab{b}})Zhao, Jin, Wang, and You]{zhao2025pab}
X.~Zhao, X.~Jin, K.~Wang, and Y.~You.
\newblock Real-time video generation with pyramid attention broadcast.
\newblock In \emph{Thirteenth International Conference on Learning
  Representations (ICLR 2025)}, 2025{\natexlab{b}}.
\newblock URL \url{https://arxiv.org/abs/2408.12588}.
\newblock arXiv:2408.12588.

\bibitem[Zheng et~al.(2024)Zheng, Yin, Xie, et~al.]{zheng2024sglang}
L.~Zheng, L.~Yin, Z.~Xie, et~al.
\newblock {SGLang}: Efficient execution of structured language model programs.
\newblock In \emph{Advances in Neural Information Processing Systems 37
  (NeurIPS 2024)}, 2024.
\newblock URL
  \url{https://proceedings.neurips.cc/paper_files/paper/2024/hash/724be4472168f31ba1c9ac630f15dec8-Abstract-Conference.html}.

\bibitem[Zhong et~al.(2024)Zhong, Liu, Chen, et~al.]{zhong2024distserve}
Y.~Zhong, S.~Liu, J.~Chen, et~al.
\newblock {DistServe}: Disaggregating prefill and decoding for
  goodput-optimized large language model serving.
\newblock In \emph{18th USENIX Symposium on Operating Systems Design and
  Implementation (OSDI 24)}, pages 193--210, 2024.
\newblock URL
  \url{https://www.usenix.org/conference/osdi24/presentation/zhong-yinmin}.

\end{thebibliography}

\end{document}